\documentclass{IEEE-Con-Sys-mag}

\usepackage{wrapfig}
\usepackage{cite}
\usepackage{algorithmic}
\usepackage{graphicx}
\usepackage[american]{circuitikz}
\usepackage{textcomp}
\usepackage{array}
\usepackage{tabu}
\usepackage{amsmath}
\usepackage{amssymb}
\usepackage{mathtools}
\usepackage[utf8]{inputenc} 
\usepackage[T1]{fontenc}    
\usepackage{hyperref}       
\usepackage{url}            
\usepackage{booktabs}       
\usepackage{amsfonts}       
\usepackage{nicefrac}       
\usepackage{microtype}      
\usepackage{color}
\usepackage{hyperref}
\usepackage{tikz}
\usetikzlibrary{arrows.meta, shapes.geometric, calc, positioning, 
decorations.pathmorphing,}

\usepackage{subcaption}
\usepackage{svg}

\definecolor{spinarrow}{RGB}{40,60,150}

\newcommand{\drawspingrid}[1]{%
\begin{tikzpicture}[baseline=(current bounding box.center)]
  \tikzset{spinnode/.style={circle, draw=black, fill=black!8, minimum size=4pt, inner sep=0pt}}
  \def\ux{0.65} \def\uy{0.00}   
  \def\vx{0.30} \def\vy{0.45}  
  \def\stubf{0.45}              
  \foreach \k in {0,...,8} {
    \pgfmathtruncatemacro{\col}{mod(\k,3)}
    \pgfmathtruncatemacro{\row}{int(\k/3)}
    \coordinate (n\k) at ({\col*\ux+\row*\vx}, {\col*\uy+\row*\vy});
  }
  \foreach \k in {0,...,8} {
    \pgfmathtruncatemacro{\col}{mod(\k,3)}
    \pgfmathtruncatemacro{\row}{int(\k/3)}
    \pgfmathtruncatemacro{\rt}{\k+1}
    \pgfmathtruncatemacro{\up}{\k+3}
    \ifnum\col<2 \draw[dashed, black!40] (n\k) -- (n\rt); \fi
    \ifnum\row<2 \draw[dashed, black!40] (n\k) -- (n\up); \fi
    \ifnum\col=0 \draw[dashed, black!40] (n\k) -- ++({-\stubf*\ux},{-\stubf*\uy}); \fi
    \ifnum\col=2 \draw[dashed, black!40] (n\k) -- ++({\stubf*\ux},{\stubf*\uy}); \fi
    \ifnum\row=0 \draw[dashed, black!40] (n\k) -- ++({-\stubf*\vx},{-\stubf*\vy}); \fi
    \ifnum\row=2 \draw[dashed, black!40] (n\k) -- ++({\stubf*\vx},{\stubf*\vy}); \fi
  }
  \foreach \s [count=\k from 0] in {#1} {
    \node[spinnode] at (n\k) {};
    \ifnum\s=1
      \draw[-{Stealth[length=1.4mm,width=1.2mm]}, thick, spinarrow] (n\k) ++(0,0.12) -- ++(0,0.26);
    \else
      \draw[-{Stealth[length=1.4mm,width=1.2mm]}, thick, spinarrow] (n\k) ++(0,0.38) -- ++(0,-0.26);
    \fi
  }
\end{tikzpicture}%
}

\pgfkeys{
  /phasegrid/.cd,
  hsep/.store in=\pgridhsep,
  vsep/.store in=\pgridvsep,
  skew/.store in=\pgridskew,
}

\newcommand{\drawphasegrid}[2][]{%
\begin{tikzpicture}[baseline=(current bounding box.center)]
  \pgfkeys{/phasegrid/.cd, hsep=0.95, vsep=0.62, skew=0.42, #1}
  \tikzset{spinnode/.style={circle, draw=black, fill=black!8, minimum size=4pt, inner sep=0pt}}
  \def\stubf{0.45}              
  \foreach \k in {0,...,8} {
    \pgfmathtruncatemacro{\col}{mod(\k,3)}
    \pgfmathtruncatemacro{\row}{int(\k/3)}
    \coordinate (n\k) at ({\col*\pgridhsep+\row*\pgridskew}, {\row*\pgridvsep});
  }
  \foreach \k in {0,...,8} {
    \pgfmathtruncatemacro{\col}{mod(\k,3)}
    \pgfmathtruncatemacro{\row}{int(\k/3)}
    \pgfmathtruncatemacro{\rt}{\k+1}
    \pgfmathtruncatemacro{\up}{\k+3}
    \ifnum\col<2 \draw[dashed, black!40] (n\k) -- (n\rt); \fi
    \ifnum\row<2 \draw[dashed, black!40] (n\k) -- (n\up); \fi
    \ifnum\col=0 \draw[dashed, black!40] (n\k) -- ++({-\stubf*\pgridhsep},0); \fi
    \ifnum\col=2 \draw[dashed, black!40] (n\k) -- ++({\stubf*\pgridhsep},0); \fi
    \ifnum\row=0 \draw[dashed, black!40] (n\k) -- ++({-\stubf*\pgridskew},{-\stubf*\pgridvsep}); \fi
    \ifnum\row=2 \draw[dashed, black!40] (n\k) -- ++({\stubf*\pgridskew},{\stubf*\pgridvsep}); \fi
  }
  \foreach \ph [count=\k from 0] in {#2} {
    \node[spinnode] at (n\k) {};
    \begin{scope}[shift={(n\k)}]
      \draw[thick, densely dashed, black] plot[domain=0:360, samples=37, smooth]
        ({\x/360*0.56*\pgridhsep-0.28*\pgridhsep},
         {0.55*\pgridvsep+0.20*\pgridvsep*sin(\x)});
      \draw[thick, spinarrow] plot[domain=0:360, samples=37, smooth]
        ({\x/360*0.56*\pgridhsep-0.28*\pgridhsep},
         {0.55*\pgridvsep+0.20*\pgridvsep*sin(\x+\ph)});
    \end{scope}
  }
\end{tikzpicture}%
}

\definecolor{inedge}{RGB}{40,60,150}
\definecolor{cutedge}{RGB}{214,32,110}

\newcommand{\partitionswap}[2][]{%
\begin{tikzpicture}[baseline=(current bounding box.center), scale=1.0,
    every node/.style={font=\small},
    site/.style={circle, fill=black!35,
                 minimum size=12pt, inner sep=0pt},
    intra/.style={inedge, thick},
    cross/.style={cutedge, very thick},
    block/.style={draw=black!50, thick}, #1]
  \coordinate (a1) at (-2.00,  0.75);
  \coordinate (a2) at (-2.00, -0.75);
  \coordinate (b1) at ( 2.00,  0.75);
  \coordinate (b2) at ( 2.00, -0.75);
  \ifnum#2=1
    \coordinate (i) at ( 0.65,  0.00);
    \draw[block] (-2.00, 0.00) ellipse[x radius=0.6, y radius=1.2];
    \draw[block] ( 1.60, 0.00) ellipse[x radius=1.3, y radius=1.2];
    \draw[white] (-1.60, 0.00) ellipse[x radius=1.3, y radius=1.5];
    \node[anchor=north] at (-2.00, -1.2) {$\tilde{\Vc}_1(k)$};
    \node[anchor=north] at ( 2.00, -1.2) {$\tilde{\Vc}_2(k)$};
  \else
    \coordinate (i) at ( -0.65, 0.00);
    \draw[block] ( 2.00, 0.00) ellipse[x radius=0.5, y radius=1.2];
    \draw[block] (-1.60, 0.00) ellipse[x radius=1.3, y radius=1.2];
    \draw[white] ( 1.60, 0.00) ellipse[x radius=1.3, y radius=1.5];
    \node[anchor=north] at (-2.00, -1.2) {$\tilde{\Vc}_1(k + 1)$};
    \node[anchor=north] at ( 2.00, -1.2) {$\tilde{\Vc}_2(k + 1)$};
  \fi
  \draw[intra] (a1) -- (a2);
  \draw[intra] (b1) -- (b2) ;
  \draw[cross] (a1) -- (b1);
  \draw[cross] (a2) -- (b2);
  \ifnum#2=1
    \draw[intra, dashed] (b1) -- (i);
    \draw[intra, dashed] (b2) -- (i);
    \draw[cross, dashed] (a1) -- (i);
    \draw[cross, dashed] (a2) -- (i);
    \foreach \n in {a1,a2} {\node[site] at (\n) {$\uparrow$};}
    \foreach \n in {b1,b2,i} {\node[site] at (\n) {$\downarrow$};}
  \else
    \draw[intra, dashed] (a1) -- (i);
    \draw[intra, dashed] (a2) -- (i);
    \draw[cross, dashed] (b1) -- (i);
    \draw[cross, dashed] (b2) -- (i);
    \foreach \n in {a1,a2,i} {\node[site] at (\n) {$\uparrow$};}
    \foreach \n in {b1,b2} {\node[site] at (\n) {$\downarrow$};}
  \fi
  \node[anchor=south] at ([yshift=3.5pt]i)  {$i_k$};
\end{tikzpicture}%
}

\newcommand{\drawPneqNP}[1][]{%
\begin{tikzpicture}[baseline=(current bounding box.center), scale=0.9,
    every node/.style={font=\footnotesize}, #1]
  \begin{scope}
    \clip (-1.8,-1.3) rectangle (1.8,2.0);
    \draw (0,2.0) ellipse[x radius=1.6, y radius=2.1];
  \end{scope}
  \draw[thick] (0,0) circle[radius=1.2];
  \draw[thick, densely dotted] (0,-0.8) ellipse[x radius=0.75, y radius=0.4];
  \node at (0,1.55)    {NP-hard};
  \node at (0,0.45)    {NP-complete};
  \node at (-0.7,-0.2) {NP};
  \node at (0,-0.8)    {P};
  \node[anchor=north] at (0,-1.5) {$\text{P}\neq\text{NP}$};
\end{tikzpicture}%
}

\newcommand{\drawPeqNP}[1][]{%
\begin{tikzpicture}[baseline=(current bounding box.center), scale=0.9,
    every node/.style={font=\footnotesize}, #1]
  \draw (-1.55,2.0) to[out=-88, in=110] (200:1.2)
    arc[start angle=200, end angle=340, radius=1.2]
    to[out=70, in=-92] (1.55,2.0);
  \draw[thick] (0,0) circle[radius=1.2];
  \node at (0,1.55) {NP-hard};
  \node[align=center] at (0,0) {$\text{P}=\text{NP}$\\$\simeq$ NP-complete};
  \node[anchor=north] at (0,-1.5) {$\text{P}=\text{NP}$};
\end{tikzpicture}%
}

\usepackage{bibunits}
\defaultbibliographystyle{IEEEtran}
\makeatletter
\newcommand{\citemain}[2][]{\leavevmode\std@@citex[#1]{#2}}
\newcommand{\sidebarbibitemfix}{\begingroup
  \renewcommand\@bibitem[1]{\item\if@filesw \immediate\write\@auxout
    {\string\bibcite{##1}{S\the\value{\@listctr}}}\fi\ignorespaces}%
}
\makeatother

\makeatletter
\newenvironment{sidebarpart}[1]{\sidebarbibitemfix
\begin{sidebar}{#1}%
\renewcommand{\thesequation}{S\arabic{sequation}}%
\renewcommand{\thestable}{S\arabic{stable}}%
\renewcommand{\thesfigure}{S\arabic{sfigure}}%

\def\@currentlabel{#1}\def\@currentlabelname{#1}%
\colorlet{csmfigurecolor}{csmsidebarfigurecolor}%
}{\end{sidebar}\endgroup}

\newenvironment{sidebarfixed}[2][tp]{\def\@dbflt##1{\@xdblfloat{##1}[#1]}%
\begin{bibunit}\begin{sidebarpart}{#2}}%
{\end{sidebarpart}\end{bibunit}}
\makeatother

\definecolor{csmsidebarcolor}{cmyk}{0.09,0.09,0,0.06}
\colorlet{csmsummarycolor}{summc}                      
\colorlet{csmsidebarinitialcolor}{black}               
\colorlet{csmsummaryinitialcolor}{summic}              
\definecolor{csmfigurecolor}{cmyk}{0.00,0.00,0.00,0.00}         
\definecolor{csmsidebarfigurecolor}{cmyk}{0.09,0.09,0,0.06} 
\definecolor{csmfigurelabelcolor}{RGB}{44,55,151}      

\mdfdefinestyle{MyFrame}{%
  linewidth=0pt, outerlinewidth=0pt, roundcorner=6pt,
  innertopmargin=\baselineskip, innerbottommargin=\baselineskip,
  innerrightmargin=12pt, innerleftmargin=12pt,
  backgroundcolor=csmsidebarcolor}
\mdfdefinestyle{Summ}{%
  linewidth=0pt, outerlinewidth=0pt, roundcorner=6pt,
  innertopmargin=\baselineskip, innerbottommargin=\baselineskip,
  innerrightmargin=12pt, innerleftmargin=12pt,
  backgroundcolor=csmsummarycolor}
\renewcommand{\summaryinitial}[1]{\lettrine[lines=2,slope=0pt,nindent=0pt]%
  {\sffamily\fontseries{bc}\selectfont\textcolor{csmsummaryinitialcolor}{#1}}{}}
\renewcommand{\sdbarinitial}[1]{\lettrine[lines=2,slope=0pt,nindent=0pt]%
  {\sffamily\fontseries{bc}\selectfont\textcolor{csmsidebarinitialcolor}{#1}}{}}

\mdfdefinestyle{CSMFigure}{%
  linewidth=0pt, outerlinewidth=0pt, roundcorner=6pt,
  innertopmargin=8pt, innerbottommargin=8pt,
  innerrightmargin=0pt, innerleftmargin=0pt,
  skipabove=0pt, skipbelow=0pt,
  backgroundcolor=csmfigurecolor}
\newlength{\figframetextwidth}
\newlength{\figframecolumnwidth}
\newenvironment{figframe}
  {\setlength{\figframetextwidth}{\textwidth}%
   \setlength{\figframecolumnwidth}{\columnwidth}%
   \begin{mdframed}[style=CSMFigure]%
   \setlength{\textwidth}{\figframetextwidth}%
   \setlength{\columnwidth}{\figframecolumnwidth}%
   \centering}
  {\end{mdframed}}

\DeclareCaptionFont{csmlabelcolor}{\color{csmfigurelabelcolor}}
\DeclareCaptionFont{csmsansmath}{\mathversion{sans}}
\makeatletter
\renewenvironment{sequation}{\bgroup
  \renewcommand{\thesequation}{\sffamily S\arabic{sequation}}%
  \let\c@equation\c@sequation\let\theequation\thesequation
  \begin{equation}}%
  {\end{equation}\egroup\ignorespacesafterend}
\def\sdbarfig#1#2{\bgroup
  \let\c@figure\c@sfigure\let\thefigure\thesfigure
  \def\theHfigure{S\arabic{sfigure}}%
  \bigskip\begin{figframe}\makebox[\linewidth]{#1}\end{figframe}%
  \caption{#2}\bigskip\egroup}
\newcounter{sbib}
\let\csm@thesbibliography\thesbibliography
\def\thesbibliography#1{\csm@thesbibliography{#1}\setcounter{enumiv}{\value{sbib}}}

\makeatother

\newcommand{\sidebarref}[1]{``\nameref{#1}''}

\newcommand{\Bc}{\mathcal{B}}

\newcommand{\Ec}{\mathcal{E}}

\newcommand{\Gc}{\mathcal{G}}

\newcommand{\Mc}{\mathcal{M}}
\newcommand{\Nc}{\mathcal{N}}

\newcommand{\Qc}{\mathcal{Q}}

\newcommand{\Vc}{\mathcal{V}}

\newcommand{\Tb}{\mathbb{T}}

\newcommand{\real}{\ensuremath{\mathbb{R}}}
\newcommand{\complex}{\ensuremath{\mathbb{C}}}

\newcommand{\intpos}{{\mathbb{N}}}

\newcommand{\norm}[1]{\left\|#1\right\|}

\newcommand{\ddfrac}[2]{\frac{d #1}{d #2}}
\newcommand{\pfrac}[2]{\frac{\partial #1}{\partial #2}}

\newcommand{\oprocendsymbol}{\hbox{$\bullet$}}
\newcommand{\oprocend}{\relax\ifmmode\else\unskip\hfill\fi\oprocendsymbol}

\newcommand{\diag}{\operatorname{diag}}

\newtheorem{csm_example}{Example}
\newtheorem{csm_definition}{Definition}

\makeatletter
\def\@begintheorem#1#2{\@IEEEtmpitemindent\itemindent\topsep12pt\rmfamily\trivlist%
    \item[\hskip\labelsep\itshape #1\ #2:]\itemindent\@IEEEtmpitemindent}
\def\@opargbegintheorem#1#2#3{\@IEEEtmpitemindent\itemindent\topsep12pt\rmfamily\trivlist%
    \item[\hskip\labelsep\itshape #1\ #2\ (#3):]\itemindent\@IEEEtmpitemindent}
\def\@IEEEproof[#1]{\par\addvspace{12pt \@plus2pt \@minus2pt}\noindent{\itshape #1:}\ \ignorespaces}
\makeatother

\jvol{}
\jnum{}
\paper{}
\jmonth{}
\jname{IEEE CONTROL SYSTEMS}
\pubyear{}

\begin{document}

\title{Computing with Coupled Oscillator Networks:\stitle{Optimization, Memory, and Modern AI Primitives}}

\author{{A}hmed Allibhoy, Fabio Pasqualetti}
\affil{Department of Electrical Engineering and Computer Science \\ University of California, Irvine.}

\maketitle

\dois{}{}

\chapterinitial{S}ynchronization of oscillations in networks of interacting components occurs in countless 
examples of systems in biology, physics, and engineering~\cite{AP-MR-JK:01, SHS:03, AA-ADG-JK-YM-CZ:08}. Often, synchronous oscillations play a crucial 
functional role in the settings they emerge from, e.g. carefully choreographed electrical activity 
of pacemaker cells in cardiac tissue regulates heart rhythm, and animal locomotion is driven by
precisely timed phase differences in oscillating nerve cells that control gait (cf.~\cite{ATW:01, LG:01}). Though these phenomena 
are numerous and diverse, they are governed by a set of universal mathematical principles, 
which capture how patterns of synchrony arise from the complex interplay between the dynamics, 
local interactions, and network interconnection structure of the system. In this article, 
we show how these mathematical models also inform the design of oscillator-based computers. 
Specifically, by emulating neural oscillations underlying human cognition and learning, coupled 
oscillator networks can perform computing operations that form the basis of artificial 
intelligence models, eventually enabling the construction of efficient and intelligent devices. 

The hegemonic paradigm of computation today is digital, with information 
encoded through binary-valued variables (called \emph{bits}), and computation carried out by 
sequentially performing logical or arithmetic operations on them. Composing these operations 
in the correct manner, combined with the ability to store and retrieve digitally encoded data,
allows digital computers to perform the extremely complex calculations necessary for the 
modern information age. Although the foundations for this approach were established as early as 
the 1940s, it nonetheless continues to be followed by nearly all modern-day computers. 
Two factors explain its persistence. First, the information encoding scheme 
and fundamental computational primitives are \emph{agnostic to hardware implementation}. 
This provides for a theory of computing abstracted away from problems of physics and 
engineering, as well as the flexibility to translate digital computing 
architectures to new devices as better technology becomes available, e.g. with 
transistors and magnetic storage replacing vacuum tubes and punch cards. 
Second, advances in semiconductor manufacturing in the
twentieth century allowed 
digital computers to continually scale up and meet the increasing computational 
demands of the last sixty years. 

However, digital computing suffers from a number of limitations that are becoming 
more acute today. Some are theoretical, such as limits on the scalability of digital 
algorithms established by computational complexity theory. 
Many of the problems 
relevant to cryptography, operations research, and data analysis are NP-hard and thus
large instances of these problems 
remain intractable on digital computers. Other limitations are physical in nature.  
Moore's law, the observation that the number of transistors on an integrated circuit doubles 
every two years, began to saturate in the 2010s, while at the same time the 
growth of AI models -- whose parameter counts now exceed the hundreds of billions -- is leading 
to concerns over their increasing energy use.

\begin{summary}
\summaryinitial{W}e explore how synchronization in coupled oscillator networks may be employed to 
perform a variety of computing tasks. 
These systems are becoming more attractive as an 
alternative to conventional digital computers due to the increasing energy demands of the 
latter, particularly for large-scale problems in machine learning and artificial 
intelligence. 
The ubiquity of coupled oscillator networks means that oscillator-based 
computing can be thought of as a universal computing paradigm encompassing systems 
with a diversity of physical realizations, e.g. as analog circuits, neurological 
systems, spintronic systems. The bridge between the dynamics of coupled oscillator 
networks and computational primitives they implement is the Ising model, which originates in 
statistical physics, and when applied to oscillator networks, explains the emergence of 
globally complex and disordered synchronization patterns from local interactions in the network. 
We discuss the application of coupled oscillator networks to three fundamental computing 
operations. First, oscillator networks as specialized hardware solvers of NP-hard combinatorial 
optimization problems, with the solution encoded in the synchrony patterns. 
Second, oscillator networks as associative memory systems, enabling storage and retrieval of data 
represented through phases of oscillations. Finally, oscillator networks as a dynamical implementation 
of attention, with synchronization providing the mechanism for the alignment of keys to queries. 
The ability of oscillator networks to implement these computational primitives paves the way for novel 
realizations of intelligent systems through a variety of physical substrates. 
\end{summary}

Despite this, usage of AI models continues to proliferate, due to their 
remarkable success on reasoning tasks, natural language processing, 
and computer vision. Such models are often inspired by neurobiological
principles. Artificial neural networks are loosely based on how biological neurons 
send and receive signals through synaptic connections, and many neural network 
architectures, such as the Hopfield model~\cite{JJH:82, JJH:84} and 
the Boltzmann machine~\cite{DHA-GEH-TJS:85},
analogize models of cognition in the human brain. 
Nevertheless, the human brain consumes a fraction 
of the energy required by computers running state-of-the-art models. One reason is the physical 
mechanism carrying out the computation: AI models run on digital hardware, 
whereas the human brain is fundamentally analog. Thus, to the extent that they
reproduce processes observed in the human brain, they are merely \emph{simulating} them and 
are unable to attain the efficiency of the human brain.

This has spurred the search for alternatives to digital computing that are 
\emph{biologically plausible}. A growing body of evidence in the neuroscientific literature 
suggests that many cognitive processes are guided by oscillatory activity in the 
human brain~\cite{FV-JPL-ER-JM:01, MB-SH-AD:10, XW:10, JF-NA:11, PF:15}. Replicating 
these oscillations in engineered systems may offer a path to building intelligent computing 
devices. While this is an active area of research, these ideas are also faithful to the 
aspirations behind many classical brain-inspired architectures such as the Hopfield model, 
or Boltzmann machines, whose physical operation, when realized through synchronizing oscillator 
networks, is closer to the processes they intend to capture in the human brain.

\begin{figure}
  \centering
\begin{figframe}
\begin{tikzpicture}[
      every node/.style={font=\scriptsize},
      hub/.style={draw, rectangle, rounded corners=3pt, align=center, inner xsep=5pt, inner ysep=3pt},
      leaf/.style={align=center, inner sep=1pt},
      src/.style={leaf, text depth=0.25ex},
      hdr/.style={align=center, font=\scriptsize\itshape},
      flow/.style={-{Stealth[length=3.5pt, width=3pt]}, shorten >=1pt}
    ]
\node[hdr]  (hw)    at (0,4.2)     {Physical realizations of oscillator networks};
\node[src]  (neuro) at (-2.5,2.85) {Neuroscience};
\node[inner sep=0pt, anchor=south] (neuroicon) at ([yshift=1pt]neuro.north) {\includegraphics[height=0.8cm]{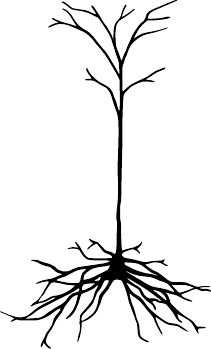}};
\node[src]  (elec)  at (0,2.85)    {Electronics};
\begin{scope}[shift={([yshift=3pt]elec.north)}, line width=0.5pt]
  \ctikzset{bipoles/length=0.55cm}
  \draw (-0.3,0) to[C] (-0.3,0.7) -- (0.3,0.7) to[L] (0.3,0) -- (-0.3,0);
\end{scope}
\node[src]  (phys)  at (2.5,2.85)  {Physics};
\begin{scope}[shift={([yshift=3pt]phys.north)}, line width=0.5pt]
  \shade[ball color=gray!12] (0,0.38) circle (0.27);
  \draw (0,0.38) circle (0.27);
  \draw[densely dashed, line width=0.3pt] (0.27,0.38) arc[start angle=0, end angle=180, x radius=0.27, y radius=0.08];
  \draw[line width=0.4pt] (0.27,0.38) arc[start angle=0, end angle=-180, x radius=0.27, y radius=0.08];
  \draw[-{Stealth[length=3.5pt, width=3pt]}] (0,0.02) -- (0,0.92);
  \draw[-{Stealth[length=3pt, width=3pt]}, line width=0.45pt]
    (0,0.76) ++(240:0.19 and 0.055) arc[start angle=240, end angle=555, x radius=0.19, y radius=0.055];
\end{scope}

\node[hub]  (con)   at (0,1.1)     {Coupled Phase Oscillator\\Networks};
\node[hub]  (ising) at (0,-0.4)    {Ising Model};
 
\node[leaf] (combo) at (-2.5,-1.7) {Combinatorial\\Optimization};
\node[inner sep=0pt, anchor=north, font=\footnotesize] (comboglyph) at ([yshift=-4pt]combo.south) {$\displaystyle\min_{s} H(s)$};
\node[leaf] (assoc) at (0,-1.7)    {Associative\\Memory};
\begin{scope}[shift={([yshift=-5pt]assoc.south)}, x=0.14cm, y=0.14cm, line width=0.3pt]
  \begin{scope}[shift={(-4.5,0)}]
    \foreach \c/\r in {0/0, 0/2, 2/2, 1/1, 0/1}
      \fill[gray!70] (\c,-\r) rectangle ++(1,-1);
    \draw[step=1] (0,0) grid (3,-3);
  \end{scope}
  \begin{scope}[shift={(1.5,0)}]
    \foreach \c/\r in {0/0, 2/0, 0/2, 2/2, 1/1}
      \fill[gray!70] (\c,-\r) rectangle ++(1,-1);
    \draw[step=1] (0,0) grid (3,-3);
  \end{scope}
  \draw[-{Stealth[length=2.5pt, width=2.5pt]}, line width=0.4pt] (-1.0,-1.5) -- (1.0,-1.5);
\end{scope}
 
\node[leaf] (atten) at (2.5,-1.7)  {Attention\\Mechanism};
\begin{scope}[shift={([yshift=-5pt]atten |- assoc.south)},
               tok/.style={inner sep=0pt, font=\scriptsize, anchor=base west}]
  \foreach \lab/\xx in {x_1/-0.55, x_2/-0.2, x_T/0.62}
    \node[tok] at (\xx-0.045,-0.15) {$\lab$};
  \foreach \lab/\xx in {y_1/-0.55, y_2/-0.2, y_T/0.62}
    \node[tok] at (\xx-0.045,-0.66) {$\lab$};
  \node[inner sep=0pt, font=\scriptsize, anchor=base] at (0.25,-0.15) {$\cdots$};
  \node[inner sep=0pt, font=\scriptsize, anchor=base] at (0.25,-0.66) {$\cdots$};
  \foreach \a in {-0.55,-0.2,0.62}
    \foreach \b in {-0.55,-0.2,0.62}
      \draw[line width=0.3pt, gray!80!black] (\a,-0.2) -- (\b,-0.5);
\end{scope}
 
\node[hdr]  (cap)   at (0,-3.3)    {Computing Capabilities};
 
\coordinate (pr) at (0,2.3);
\draw[rounded corners=4pt] (neuro.south) |- (pr);
\draw[rounded corners=4pt] (phys.south)  |- (pr);
\draw (elec.south) -- (pr);
\draw[flow] (pr) -- node[right, font=\scriptsize\itshape, inner sep=3pt] {phase reduction} (con.north);
\draw[flow] (con)   -- (ising);
\draw[flow] (ising) -- (combo);
\draw[flow] (ising) -- (assoc);
\draw[flow] (ising) -- (atten);
\end{tikzpicture}
\end{figframe}
\caption{Overview of relationships between various concepts introduced 
  in this survey. Phase reduction theory connects mathematical descriptions 
  of a variety of physical realizations of oscillator networks to 
  universal coupled phase oscillator models. The Ising model 
  provides a bridge between coupled oscillator networks and 
  their computing capabilities.}
\label{fig:concepts}
\vspace{-1.5ex}
\end{figure}
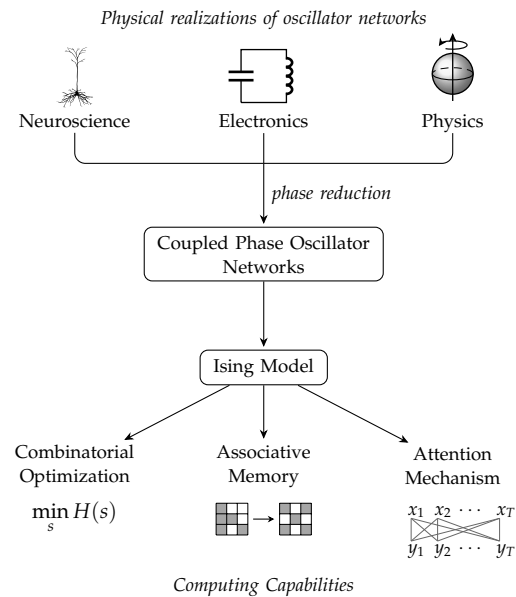

\begin{sidebarfixed}{Oscillator-Based Computing: Then and Now}

  \sdbarinitial{W}hile digital computers based on the von Neumann architecture are the 
  predominant approach to computing today, analog computing has an equally 
  long history. In fact many of the predecessors to modern-day computers 
  were special-purpose analog devices, for example the tide-predicting 
  machine built by Sir William Thomson which used rotational components 
  to represent the phases of lunar and solar tidal cycles~\cite{WT:81}. 
  Even in the early days of digital computing it was recognized that 
  oscillator-based computers may offer advantages in terms of speed 
  and energy efficiency, with von Neumann himself proposing to 
  replace vacuum tubes with nonlinear electrical 
  components whose oscillatory phases represent bits~\cite{JVN:57, RLW:59}. 
  An example of such a device is the parametron which was used to 
  construct the MUSASINO-1 in 1957~\cite{SM-KT:59}. 

  \phantom{\cite{NS-AP-MC-MB-XL-NC-HP-DGS-VN-AR-SD:14-iedm}\cite{MKB-AM-DST-BHC-SJ-NS:20} \cite{KS-ML-SK-OS-SK-PC-DS:17, SR-MK-NY-MP:24}
  \cite{TC-RKD-AE-PKM-AH-AAA-PD-BGM-AR-JA:16, JT-MR-FAA-ST-GK-DQ-PB-VC-KY-AF-HK-SY-MDS-JG:17, AK-AL-NB-AAA-RK-JA:24}
  \cite{ZW-AM-KW-RLB-YY:13, PLM-AM-YH-RH-CL-ST-TI-HT-SU-KA-RLB-MMF-HM-YY:16}
  \cite{JCC-MCAY-JS:17}}

  \sdbarfig{
    \includegraphics[height=0.4\linewidth]{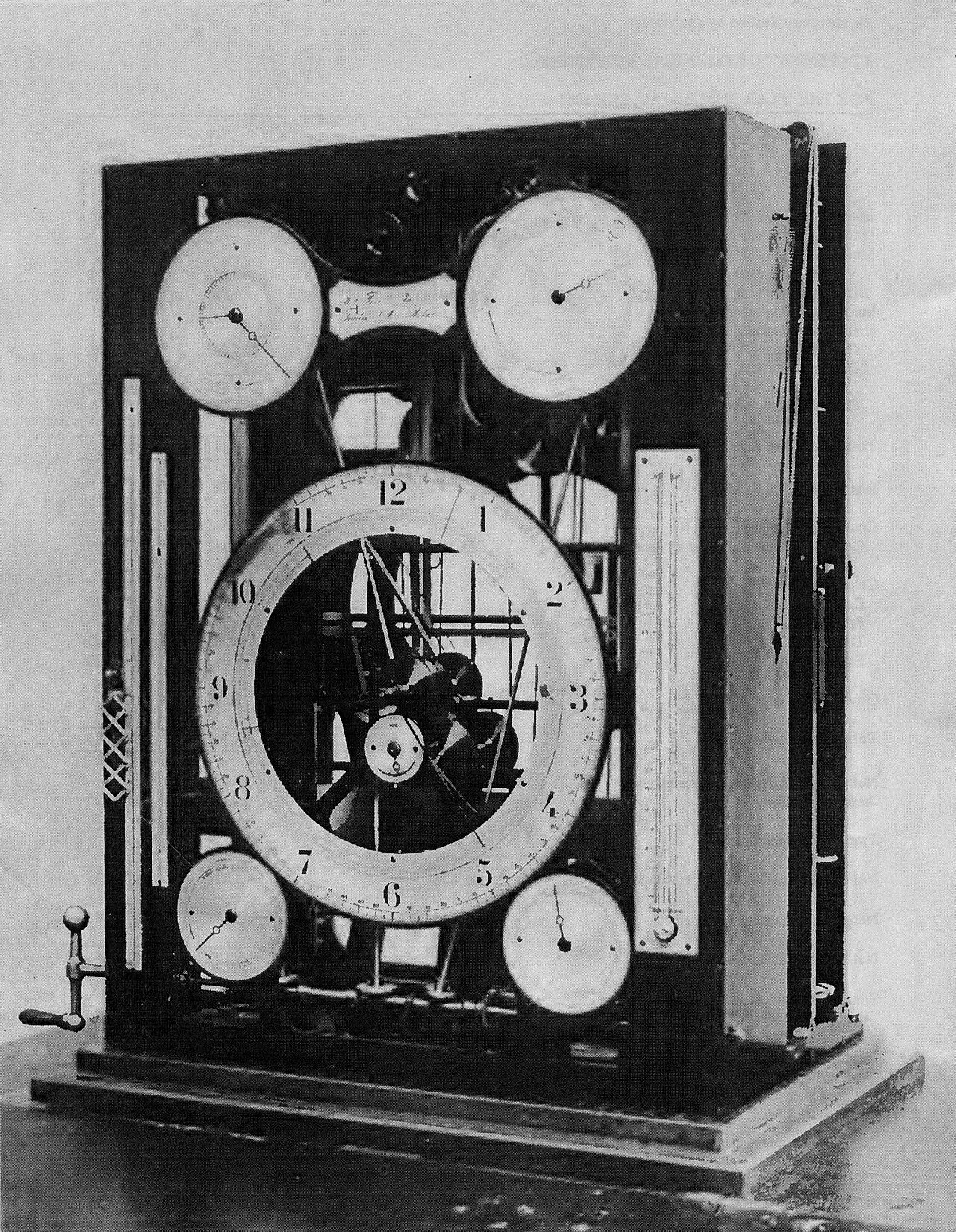}
    \includegraphics[height=0.4\linewidth]{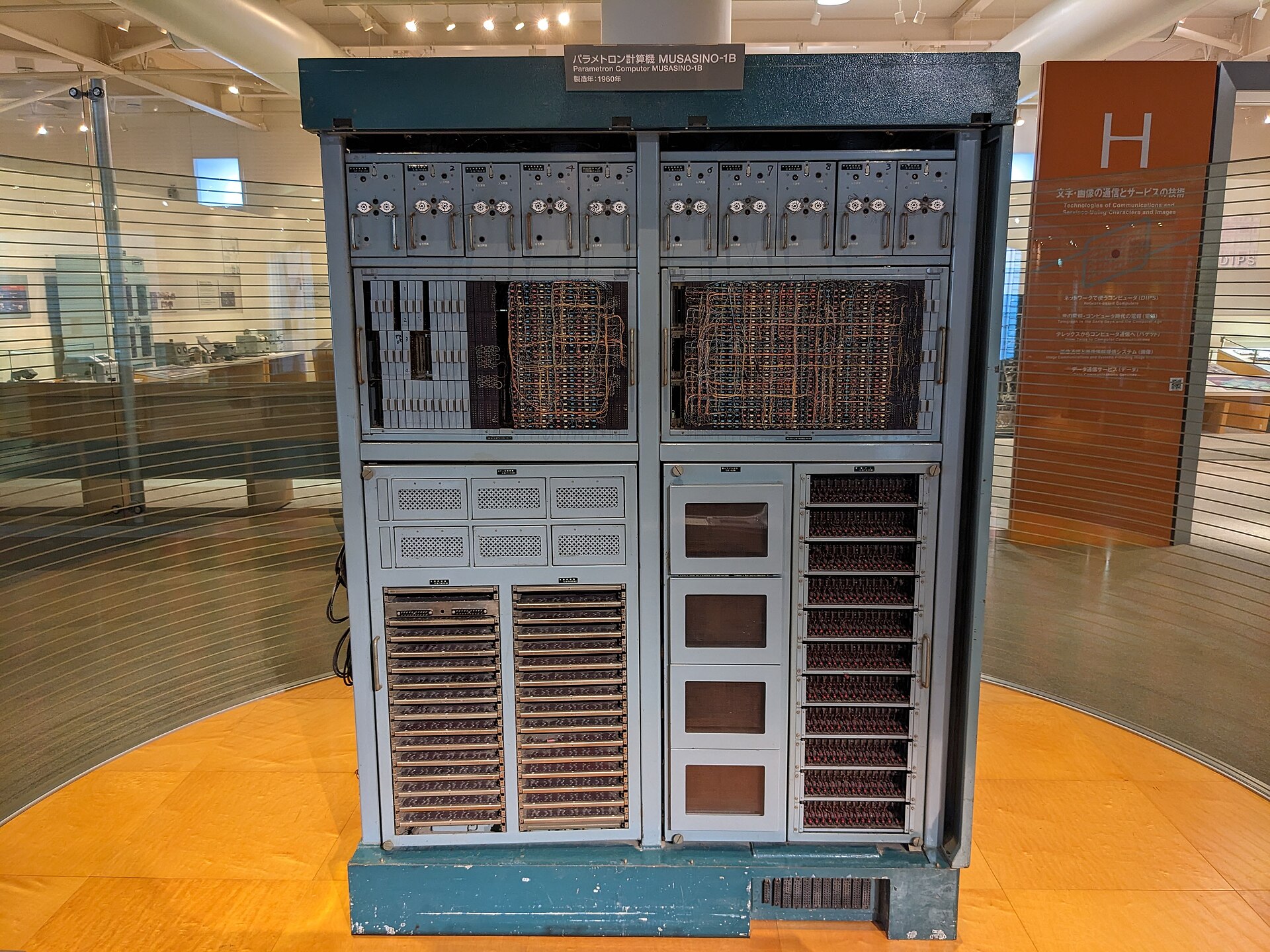}
  }{Classical examples of oscillator-based computers. (Left) Thomson's tide-predicting machine (Right) MUSASINO-1 parametron computer}

  \sdbarfig{
    \includegraphics[height=0.4\linewidth]{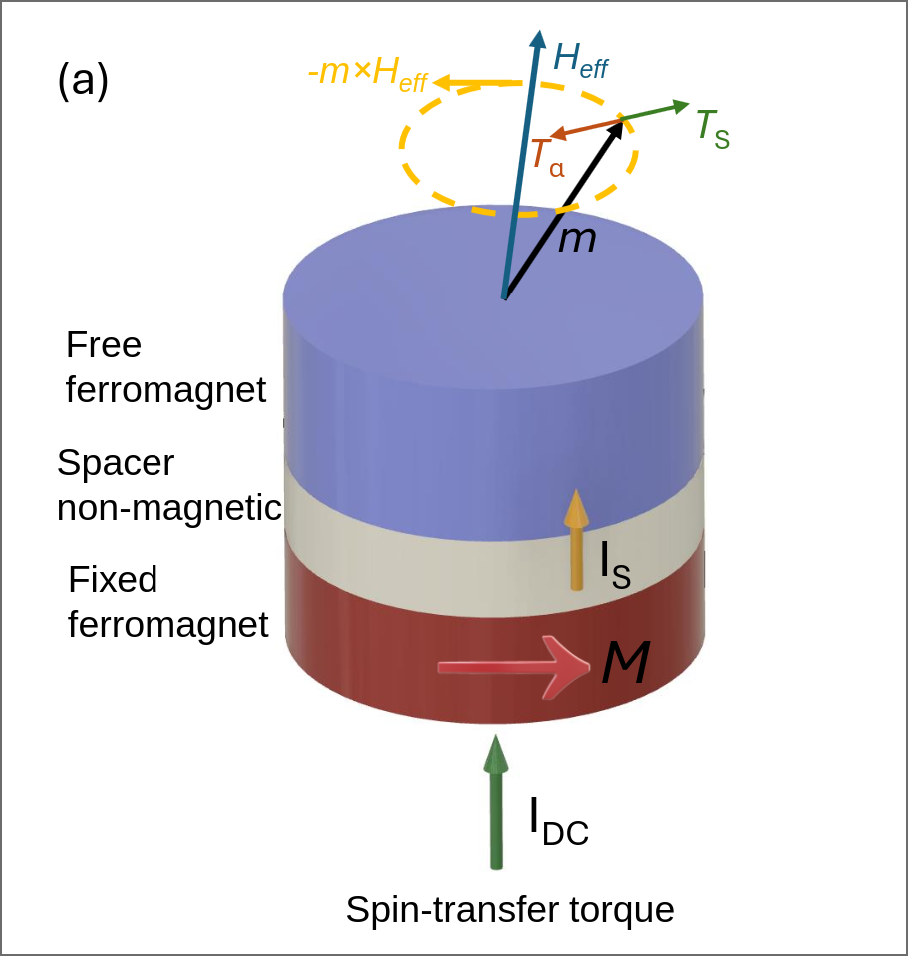}
    \includegraphics[height=0.4\linewidth]{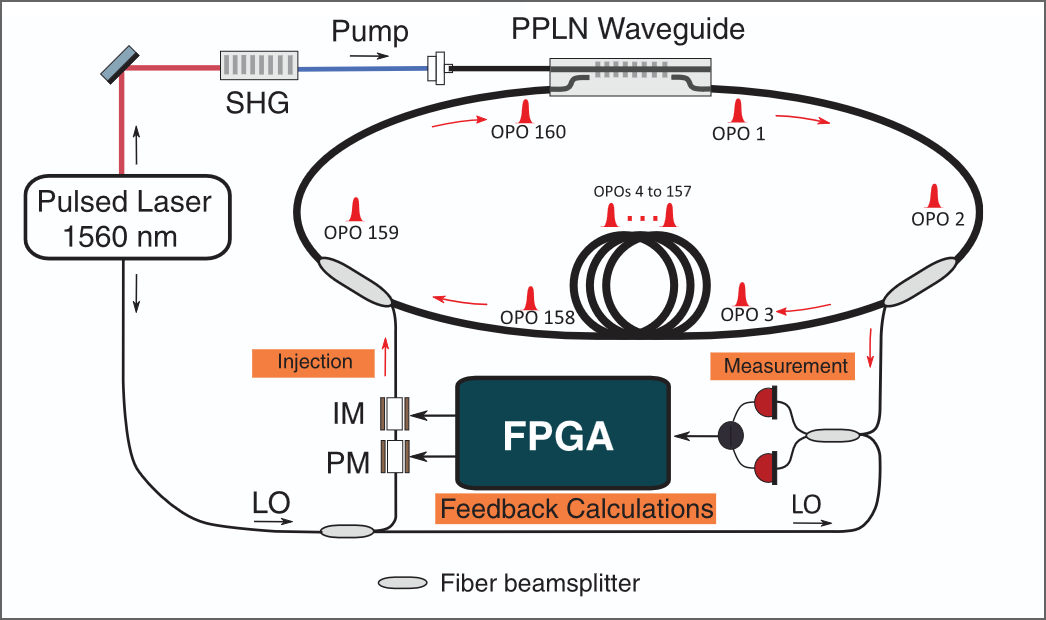}
  }{Modern examples of oscillator-based comptuers 
  (Left) Spin torque oscillator~\cite{AK-AL-NB-AAA-RK-JA:24} 
  (Right) Computer based on optical parametric oscillators~\cite{PLM-AM-YH-RH-CL-ST-TI-HT-SU-KA-RLB-MMF-HM-YY:16} }

  In recent years there has been a resurgence of interest in oscillator-based 
  computers to bypass the limitations of digital computers. Such devices are capable 
  of implementing computational primitives such as solving combinatorial 
  optimization problems, or implementing associative memory, and are
  realized through a variety of physical mechanisms, including

  \begin{enumerate}
    \item \textbf{Conventional electronics}: analog circuits such as LC tank circuits \cite{NS-AP-MC-MB-XL-NC-HP-DGS-VN-AR-SD:14-iedm} or relaxation oscillators \cite{MKB-AM-DST-BHC-SJ-NS:20}
    \item \textbf{Superconducting Josephson junctions}: devices using superconducting circuit elements with oscillating voltages \cite{KS-ML-SK-OS-SK-PC-DS:17, SR-MK-NY-MP:24}
    \item \textbf{Spintronic oscillators}: devices which utilize the precession of magnetic moments in materials, such as spin-torque oscillators and spin Hall nano-oscillators \cite{TC-RKD-AE-PKM-AH-AAA-PD-BGM-AR-JA:16, JT-MR-FAA-ST-GK-DQ-PB-VC-KY-AF-HK-SY-MDS-JG:17, AK-AL-NB-AAA-RK-JA:24}
    \item \textbf{Optoelectronic oscillators}: devices such as degenerate optical parametric oscillators (DOPOs), where oscillations are generated using coupled lasers \cite{ZW-AM-KW-RLB-YY:13, PLM-AM-YH-RH-CL-ST-TI-HT-SU-KA-RLB-MMF-HM-YY:16}
    \item \textbf{Electromechanical oscillators}: devices built from oscillating microelectromechanical (MEMS) or nanoelectromechanical (NEMS) systems \cite{JCC-MCAY-JS:17}
  \end{enumerate}

  \putbib[refs]
\end{sidebarfixed}

Here, we review a paradigm for computing using oscillator networks, where the information 
encoding scheme is analog, and utilizes the phase of an oscillation.
The mechanism that drives 
computation is the dynamics of the network system, along which phases evolve 
continuously in time. 
The ubiquity of oscillator networks in both naturally occurring and 
engineered settings means that, similar to digital computing, these devices may be 
realized through a number of physical substrates~\cite{GC-AR-SD-WP:18-iscas, AR-AP-GHS-VN-GC-MJ-WP-SD:19, GC-WP:20}. 
Oscillator networks offer a number of advantages 
over conventional digital computers. First, certain realizations of 
analog computers achieve 
greater energy efficiency than digital computers. 
Second, because information is analog and evolves 
continuously in time, rather than digital and taking place in discrete increments, these devices 
are not subject to the same theoretical limitations imposed by computational complexity 
theory on digital computers.

This is not to say that oscillator-based 
computing comes without tradeoffs, but rather to motivate the need for a new 
\emph{theory of oscillator-based computing} to systematically explore them. 
Such a theory would consist of the following elements:
\begin{enumerate}
  \item\label{item:framework} An abstract framework for oscillator-based computing expressed independently of 
  any physical realization, within which one can develop 
  computing architectures and reason about their capabilities and limitations.
  \item A set of mathematical tools to relate the physics of
  real-world oscillatory systems to a simplified and universal representation in terms 
  of the phases alone, allowing computing architectures developed within the aforementioned
  framework to be implemented on that particular platform. 
\end{enumerate}
In this survey we focus
 mainly on the former, though the latter 
is equally interesting and crucial to the practical implementation of 
oscillator-based computers (and we believe it can be addressed using 
phase reduction theory, though many open questions 
remain, see Sidebar: \sidebarref{sidebar:phase-reduction}). 

Our approach, illustrated in Figure~\ref{fig:concepts}, 
exploits the fact that many fundamental operations 
in computing are expressible in terms of minimizing an energy function 
related to the Ising Hamiltonian in models of spin glasses. Since similar
energy functions also characterize the synchrony patterns 
exhibited by coupled oscillator networks, physical realizations of these networks 
be used as analog computers implementing those operations. 
We focus on three computing primitives: (1) coupled oscillator 
networks as solvers of combinatorial optimization problems, (2) 
coupled oscillator networks as associative memory systems, which 
in turn can be used for pattern recognition and generative modeling, 
and (3) oscillator-based implementations of the attention mechanism.


\section{The Ising Model}

We begin by reviewing the Ising model from a mathematical perspective (we refer the reader 
to~\cite{KB-AY:86, MM-GP-MAV:87, MM-AM:09, MT:11} for an introduction to spin-glass models, 
including the Ising model, and their applications to physics). Though it originates in statistical mechanics, 
and was traditionally proposed to explain the magnetic properties of materials, 
it has seen wide-ranging applications in chemistry, materials science, 
neuroscience, and machine learning, and  is better understood as a general framework to 
model systems with both cooperative 
and competitive local interactions and the global order (or disorder) that emerges 
from them. The connection between the Ising model and certain fundamental operations 
in computing enables it to serve as a bridge between coupled oscillator networks and 
the computational primitives they implement. 

\begin{figure}
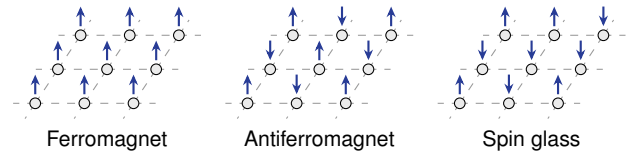

  \centering
  \begin{figframe}
  \begin{subfigure}{0.32\linewidth}
    \centering
    \drawspingrid{1,1,1,1,1,1,1,1,1}
    \caption*{Ferromagnet}
  \end{subfigure}
  \hfill
  \begin{subfigure}{0.32\linewidth}
    \centering
    \drawspingrid{1,-1,1,-1,1,-1,1,-1,1}
    \caption*{Antiferromagnet}
  \end{subfigure}
  \hfill
  \begin{subfigure}{0.32\linewidth}
    \centering
    \drawspingrid{1,-1,1,-1,-1,-1,1,1,-1}
    \caption*{Spin glass}
  \end{subfigure}
  \end{figframe}
  \caption{Types of ordering of spin configurations. Spins are aligned in
  the same direction in a ferromagnet, aligned in alternating directions
  in an antiferromagnet, and disordered in a spin glass.}
  \label{fig:order}
\end{figure}

Let $\Gc = (\Vc, \Ec)$ be a graph with $N$ nodes and signed adjacency matrix 
$J \in \real^{N \times N}$. Here, nodes represent individual agents, 
and edges represent the interactions between them. For each node, 
$i \in \{1, 2, \dots, N \}$, we assign a \emph{spin} denoted by $s_i \in \{-1, 1\}$. 
The vector of spins $s = (s_1, \dots, s_N)$
is called a \emph{spin configuration}, to which we associate an energy 
given by the value of the \emph{Ising Hamiltonian},
\begin{equation}
  \label{eq:ising-hamiltonian}
  H(s) = -\frac{1}{2}\sum_{i=1}^{N}\sum_{j=1}^{N}J_{ij}s_is_j - \sum_{i=1}^{N}h_is_i,
\end{equation}
where $h \in \real^N$ is called the \emph{external field}. 
We seek to understand the \emph{ordering} of spin configurations, 
i.e. the regularity in the alignment of spins in the network 
(cf. Figure \ref{fig:order} for examples of types of ordering of 
spin configurations). Systems
having spin configurations without any structure or regularity are 
called \emph{disordered}. 
As we show next, the ordering of observed states is determined
by the properties of low energy states. 

In typical applications, the state occupied by the system is assigned
randomly according to 
the \emph{Boltzmann distribution}, 
\begin{equation}
  \label{eq:boltzmann}
  p(s) = \frac{1}{Z_\beta}\exp(-\beta H(s)).
\end{equation}
The parameter $\beta \geq 0$ is called the \emph{inverse temperature}, and~$Z_\beta$ is 
a normalization constant $Z_\beta = \sum_{s \in \{-1, 1\}}e^{-\beta H(s)}$, 
called the \emph{partition function}. 
The likelihood of observing the 
state~$s$ varies inversely with its energy $H(s)$, biasing 
the system toward low energy states. The parameter $\beta$
determines the level of bias: in the high-temperature 
limit ($\beta \to 0^+$), the Boltzmann distribution converges to the uniform 
distribution, and the spins are almost always disordered. In the low temperature limit 
($\beta \to \infty$), the distribution concentrates around the  
global minimizers of~\eqref{eq:ising-hamiltonian}, called \emph{ground states}. 

To determine the properties of low energy states, 
it is useful to consider the \emph{local interaction energies},~$-J_{ij}s_is_j$, 
between individual pairs of nodes. In the absence of an external 
field, the Ising Hamiltonian is simply the sum of the local 
interaction energies. There are two types of interactions,
distinguished by the sign of the corresponding edge: if $J_{ij} > 0$, 
the interaction along $(i, j)$ is called \emph{ferromagnetic}, in 
which case the local energy is minimized whenever $s_i = s_j$. 
On the other hand,  if~$J_{ij} < 0$, 
the interaction is called \emph{antiferromagnetic} 
and the local energy
is instead minimized when~$s_i \neq s_j$. The external field adds 
an additional bias to each spin: if $h_i > 0$ (resp. $h_i < 0$) then the 
$i$th node is biased toward $s_i = 1$ (resp. $s_i = -1$). 

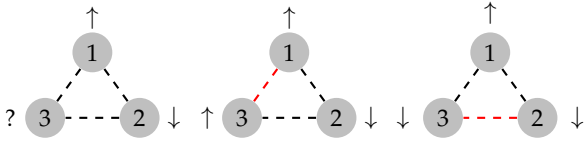
\begin{figure}
    \centering
    \begin{figframe}
    \begin{subfigure}{0.3\linewidth}
        \centering
        \begin{tikzpicture}[node distance=13pt]
            \tikzstyle{spinsite}=[circle,fill=black!25,minimum size=15pt,inner sep=0pt, node distance=60pt]
    
            \node (n1) at (0, 0) [spinsite] {$1$};
            \node (n2) at (1.25*0.5, -0.866) [spinsite] {$2$};
            \node (n3) at (-1.25*0.5, -0.866) [spinsite] {$3$};
    
            \draw[thick, dashed] (n1) -- (n2);
            \draw[thick, dashed] (n2) -- (n3);
            \draw[thick, dashed] (n3) -- (n1);
    
            \node[above of=n1] {$\uparrow$};
            \node[right of=n2] {$\downarrow$};
            \node[left of=n3] {?};
        \end{tikzpicture}
    \end{subfigure}
    \begin{subfigure}{0.3\linewidth}
        \centering
        \begin{tikzpicture}[node distance=13pt]
            \tikzstyle{spinsite}=[circle,fill=black!25,minimum size=15pt,inner sep=0pt, node distance=60pt]
    
            \node (n1) at (0, 0) [spinsite] {$1$};
            \node (n2) at (1.25*0.5, -0.866) [spinsite] {$2$};
            \node (n3) at (-1.25*0.5, -0.866) [spinsite] {$3$};
    
            \draw[thick, dashed] (n1) -- (n2);
            \draw[thick, dashed] (n2) -- (n3);
            \draw[thick, dashed, color=red] (n3) -- (n1);
    
            \node[above of=n1] {$\uparrow$};
            \node[right of=n2] {$\downarrow$};
            \node[left of=n3] {$\uparrow$};
        \end{tikzpicture}
    \end{subfigure}
    \begin{subfigure}{0.3\linewidth}
        \centering
        \begin{tikzpicture}[node distance=15pt]
            \tikzstyle{spinsite}=[circle,fill=black!25,minimum size=15pt,inner sep=0pt, node distance=60pt]
    
            \node (n1) at (0, 0) [spinsite] {$1$};
            \node (n2) at (1.25*0.5, -0.866) [spinsite] {$2$};
            \node (n3) at (-1.25*0.5, -0.866) [spinsite] {$3$};
    
            \draw[thick, dashed] (n1) -- (n2);
            \draw[thick, dashed, color=red] (n2) -- (n3);
            \draw[thick, dashed] (n3) -- (n1);
    
            \node[above of=n1] {$\uparrow$};
            \node[right of=n2] {$\downarrow$};
            \node[left of=n3] {$\downarrow$};
        \end{tikzpicture}
    \end{subfigure}
   
    \end{figframe}
    \caption{Example of a frustrated network with antiferromagnetic interactions 
    (depicted by dashed lines) between nodes, where for any spin configuration 
    there exists a pair of adjacent nodes with aligned spins. 
    }
    \label{fig:frustration}
\end{figure}

In general, the relationship between the global regularity of spin configurations 
and local interactions in the network is nontrivial. An important 
property of many networks is \emph{frustration}, where no spin configuration 
exists that simultaneously minimizes all the local interaction energies 
(cf. Figure~\ref{fig:frustration}). One observes a  
variety of complex behaviors in these systems, including distinct phases with abrupt 
temperature-dependent phase transitions, and the emergence of spontaneous order. 
The Ising model is used as a simplified model of a
\emph{spin glass},
a class of 
magnetic materials exhibiting disorder or irregularity in the alignment 
of spins even at low temperatures.

The Ising model can be generalized to allow spins
represented as unit vectors on the $(d - 1)$-sphere,
$z_i \in S^{d - 1}$ where $S^{d - 1}= \{z_i \in \real^d \mid \norm{z_i} = 1 \}$  
(we denote vector-valued spins by the letter $z$ to avoid confusion with 
scalar-valued spins). 
This model is called the $d$-vector model, and the Hamiltonian is 
\begin{equation}
  \label{eq:d-vector-hamiltonian}
  H(z) = -\frac{1}{2}\sum_{i=1}^N\sum_{j=1}^N J_{ij}z^\top_i z_j - \sum_{i=1}^{N}h^\top_iz_i.
\end{equation}
When $d = 1$, \eqref{eq:d-vector-hamiltonian} reduces to the 
classical Ising model. Higher-dimensional models exhibit similarly complex behavior 
and, as we show later, also serve an important role in understanding the 
attention mechanism. 

\subsection{The Ising Model and Coupled Oscillator Networks}
\begin{figure}
  \begin{figframe}
  \begin{subfigure}[c]{0.5\linewidth}
    \hspace{-6.0ex}
    \includegraphics[width=0.7\textwidth]{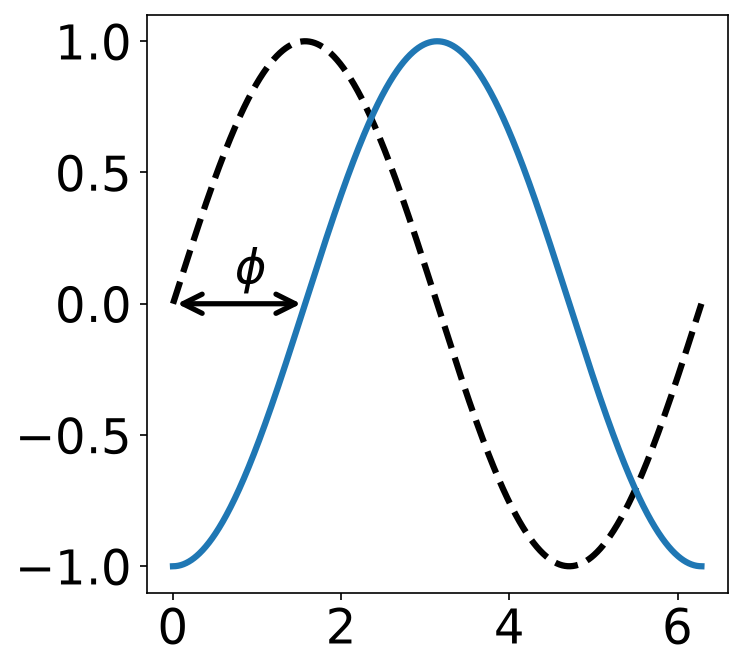}
  \end{subfigure}
  \hspace{-14.0ex}\begin{subfigure}[c]{0.4\linewidth}
    \vspace{3.0ex}
    \drawphasegrid[hsep=1.3, vsep=1.0]{0,140,250,40,310,90,200,20,160}
  \end{subfigure}
  \end{figframe}
  \caption{(Left) The blue waveform has phase shift $\phi$ relative to the reference waveform depicted in black. 
  (Right) Network where each node is associated to a phase shifted oscillation, depicted in blue, relative 
  to the uniform reference in black. }
  \label{fig:phase-network}
\end{figure}

We now discuss the relationship between the Ising model 
and synchrony patterns exhibited by coupled oscillator networks.
Instead of a spin, we associate each node $i \in \{1, 2, \dots, N \}$ 
to a \emph{phase} 
$\phi_i \in [0, 2\pi]$, which denotes the phase shift of 
its oscillation relative to a fixed reference 
(cf. Figure \ref{fig:phase-network}).  
We say that two nodes are \emph{phase synchronized}
if they have the same phase. 
Phase synchronization of 
oscillators in a network
admits a similar mathematical description as the alignment 
of binary-valued spins in the Ising model. This connection 
may be exploited to describe synchronization patterns via 
energy functions, analogous 
to~\eqref{eq:ising-hamiltonian}, 
that encode the tendency of the oscillations of a pair of 
nodes to synchronize or desynchronize due to their local interaction 
in the network. 

\begin{sidebarfixed}{The Kuramoto Model of Synchronization}
  \label{sidebar:kuramoto}
  \sdbarinitial{S}ynchronization of oscillatory systems is observed in numerous disciplines
  including physics (e.g. microwave oscillators and superconducting 
  Josephson junctions), chemistry (oscillating chemical reactions), 
  neuroscience (e.g. activity of individual neurons and brain networks),
  engineering (e.g. electric power distribution networks), 
  ecology (e.g. synchronous flashing of fireflies or chirping crickets), 
  and social networks (e.g. opinion dynamics), see e.g. 
  \cite{SS:00-prd, JAA-LLB-CJPV-FR-RS:05, SG-AC-SR:14, FD-FB:14, FR-TP-PJ-JK:16}
  and references therein. 
  
  Despite the sheer breadth of these examples, they 
  are often studied within the common framework of coupled 
  phase dynamics. In this setting, one assumes that the phase of each 
  oscillator, in the absence of coupling, evolves at a constant rate 
  according to a natural frequency $\dot{\phi}_i = \omega_i$. Interactions 
  between nodes are modeled with a coupling function $\Gamma(\phi_j - \phi_i)$,  
  resulting in a system generically of the form 
  \begin{sequation}
    \label{seq:generalized-kuramoto}
    \dot{\phi}_i = \omega_i + \sum_{j=1}^{N}J_{ij}\Gamma(\theta_j - \theta_i)
  \end{sequation}
  %
  Since $\Gamma$ is periodic, it can be expanded as a Fourier series 
  $\Gamma(\psi) = \sum_{k=-\infty}^{\infty}c_ke^{ik\psi}$. The \emph{Kuramoto model} is 
  the special case where the series contains only the first odd 
  harmonic~$\Gamma(\psi) = \sin(\psi)$, resulting in 
  \begin{sequation}
    \label{seq:kuramoto-sidebar}
    \dot{\phi}_i = \omega_i + \sum_{j=1}^{N}J_{ij}\sin(\theta_j - \theta_i).
  \end{sequation}
  Originally introduced in~\cite{YK:75}, initial work investigated the 
  setting where the natural frequencies 
  are drawn from the standard normal distribution $\omega_i \sim \Nc(0, 1)$,
  and the coupling is all-to-all:~$J_{ij}=\frac{K}{N}$ for all pairs $(i, j)$. 
  %

  %
  %

  Since its introduction, a multitude of variations of the Kuramoto model have been studied, 
  including 
  \begin{itemize}
    \item Kuramoto models with inertia 
    \[ m\ddot{\theta} + b\dot{\theta} = \omega_i + \sum_{j=1}^{N}J_{ij}\sin(\theta_j - \theta_i)  \]
    \item Kuramoto models with delays 
    \[ \dot{\theta} = \omega_i + \sum_{j=1}^{N}J_{ij}\sin(\theta_j(t - \tau_{ij}) - \theta_i)  \]
    \item Kuramoto models with biased couplings 
    \[ \dot{\theta} = \omega_i + \sum_{j=1}^{N}J_{ij}\sin(\theta_j - \theta_i - \psi_{ij})  \]
  \end{itemize}
  
  \sdbarfig{
    \begin{tikzpicture}[scale=0.95]
      \draw[thick] (0,0) circle (1.5);
      \draw[dashed] (0,0) -- (47:1.5);
      \draw[dashed] (0,0) -- (1.5,0);
      \draw (0.5,0) arc[start angle=0, end angle=47, radius=0.5];
      \node at (23.5:0.8) {$\phi_i$};
      \draw[thick, -{Stealth}] (25:1.9) arc[start angle=25, end angle=75, radius=1.9];
      \node at (50:2.3) {$\omega_i$};
      \filldraw[fill=spinarrow, draw=black, thick] (47:1.5) circle (0.15);
    \end{tikzpicture}%
    \hspace{2em}%
    \begin{tikzpicture}[scale=0.95]
      \draw[thick] (0,0) circle (1.5);
      \draw[thick, -{Stealth}] (25:1.9) arc[start angle=25, end angle=75, radius=1.9];
      \foreach \a in {12,63,118,160,203,231,274,298,341}
        \draw (47:1.5) -- (\a:1.5);
      \foreach \a in {12,63,118,160,203,231,274,298,341}
        \filldraw[fill=white, draw=black, thick] (\a:1.5) circle (0.15);
      \filldraw[fill=spinarrow, draw=black, thick] (47:1.5) circle (0.15);
    \end{tikzpicture}
  }{(Left) Single oscillator with phase $\phi_i$ changing at constant rate $\omega_i$ 
  (Right) Coupled oscillator network where evolution of $\phi_i$ now depends on the interactions 
  with other oscillators.}

  The appeal of the Kuramoto model is both its many generalizations having a rich 
  repertoire of behaviors, many of which are useful for computing applications, but also 
  its \emph{universality}. Since these models are expressed only in terms of the phases 
  of oscillations, abstracting away the physical details of any particular instance of 
  an oscillator network, they serve as universal models for computing with oscillatory systems. 
  \putbib[refs]
\end{sidebarfixed}

We begin by extending models of spin systems to the space of oscillatory 
phases. 
The vector of phases $\phi = (\phi_1, \dots, \phi_N)$ is called a 
\emph{phase configuration}, and $\Tb^N := [0, 2\pi]^N$ consists of 
all possible phase configurations of the network. 
The direct analog of the Ising model on $\Tb^N$
uses the energy function $E:\Tb^N \to \real$ defined as 
\begin{equation}
  \label{eq:XY-energy}
  E(\phi) = -\frac{1}{2}\sum_{i=1}^N\sum_{j=1}^NJ_{ij}\cos(\phi_j - \phi_i) - \sum_{i=1}^{N}h_i\cos(\phi_i). 
\end{equation}
In statistical mechanics, the model of spin systems with this energy 
function is called the \emph{XY-model}, which is equivalent 
to the $d=2$ case of the $d$-vector model~\eqref{eq:d-vector-hamiltonian} 
(cf.~\cite{JA-LB-CP-FR-RS:05} for details on the relationship between the
XY-model and coupled phase oscillators).
The local interaction energy along the edge $(i, j)$ is $-J_{ij}\cos(\phi_j - \phi_i)$, 
and behaves similarly to the corresponding 
term in the Ising model: when~$J_{ij} > 0$, it is minimized when $\phi_i = \phi_j$ i.e. when $i$ 
and $j$ are phase synchronized, and when~$J_{ij} < 0$, it is minimized 
when $\phi_j - \phi_i = \pi$, i.e. when the oscillations are exactly
one half cycle out of phase. 

Unlike a discrete spin, the phase of an oscillation varies continuously, 
and we can compute the ground states of the energy~\eqref{eq:XY-energy}
using a continuous-time dynamical system. When $h_i = 0$ 
the gradient dynamics,~$\dot{\phi} = -\nabla_\phi E(\phi)$, is
a special case of the celebrated Kuramoto model 
(see Sidebar: \sidebarref{sidebar:kuramoto}),
\begin{equation}
  \label{eq:kuramoto}
    \dot{\phi}_i = \sum_{j=1}^{N}J_{ij}\sin(\phi_j - \phi_i)
\end{equation}
We may also consider gradient dynamics that are noisy, 
\begin{equation}
  \label{eq:stochastic-kuramoto}
  d\phi_i = \sum_{j=1}^{N}J_{ij}\sin(\phi_j - \phi_i)dt + \sqrt{2\beta^{-1}}dw_i
\end{equation}
where $w_i$ is a Wiener process and $\beta$ is the inverse temperature. 
The density of phases at time $t$, given by $\rho_t(\phi)$,
satisfies $\rho_t \to \rho$  as $t\to\infty$ where $\rho$
is the Boltzmann distribution 
\[ \rho(\theta) = \frac{1}{Z_\beta}\exp(-\beta E(\theta)). \]
Coupled oscillator networks described by~\eqref{eq:kuramoto} 
or~\eqref{eq:stochastic-kuramoto} 
exhibit a range of complex behaviors, 
including frustration and disorder analogous to 
the Ising model. See, e.g.,~\cite{HD:87, HD:92, LB-CP-JR:93} and references 
therein for further details of disordered synchrony patterns of 
the Kuramoto model and their connections to spin glasses. 


Gradient systems can also identify the ground 
states of higher dimensional spin models~\eqref{eq:d-vector-hamiltonian}, 
however the standard gradient flow 
must be replaced by a projected gradient flow in order to enforce the 
constraint that $z_i \in S^{d - 1}$. In this case, the dynamics are
\begin{equation}
  \label{eq:kuramoto-sphere}
  \dot{z}_i = (I - z_iz^\top_i)\sum_{j=1}^{N}J_{ij}z_j
\end{equation}
where $(I - z_iz^\top_i)$ is the orthogonal projection onto the 
tangent space $T_{z_i}S^{d - 1}$. This system
is a generalization of the Kuramoto model and is well-studied in 
the control community to model swarming behavior and consensus in 
multi-agent systems on the 
$(d - 1)$-sphere~\cite{ROS:06-cdc, JM-JT-JG:18, SC-MG-EO:19, ML-RM-SHS:21, JM-DP-LM-JG:21}, 
and is also related to models that have been studied in the physics community of 
synchronization in non-Abelian spaces and quantum mechanical systems~\cite{MAL:09, MAL:10}. 
In the special case where $d=2$, 
the system reduces to~\eqref{eq:kuramoto}. 

Our interest in systems of the form~\eqref{eq:kuramoto} 
and~\eqref{eq:stochastic-kuramoto}
is in their connection to computing. Specifically 
\begin{itemize}
  \item \textbf{Many computing operations can be reduced to
  identifying the ground states of an energy function,
  opening the door to implementation via a physical process that minimizes the energy.}
  Such systems, 
  called \emph{energy-based dynamical models}, are well-studied and have deep connections to computational 
  neuroscience, machine learning, and physical computing~\cite{YL-SC-RH-MAR-FJH:06, AM-FB-DK-AM:26}. 
  \item \textbf{Phase oscillator models are universal and admit a diversity of physical realizations,}
  This is because they are expressed entirely in terms 
  of oscillatory phases, abstracting away physical details of any particular realization. 
  The dynamics of a specific instance of an oscillatory system can be reduced to the models 
  considered in this paper using \emph{phase reduction theory}, see Sidebar: \nameref{sidebar:phase-reduction}. 
\end{itemize}

To design oscillator-based implementations of computational primitives, 
we proceed by approximating the Ising Hamiltonian
with an energy function similar to~\eqref{eq:XY-energy}, 
so the solution can then be computed with the corresponding gradient flow. 
In certain applications 
we will need to add additional penalty terms or higher order 
harmonics to~\eqref{eq:XY-energy} in order to transform the problem 
into a form suitable for implementation on coupled oscillator networks. 
However, once a suitable energy has been identified, if the gradient 
system can be realized through the dynamics of a physical device, 
then it can be used as an analog computer 
implementing that computational primitive.


\begin{bibunit}
\begin{sidebarpart}{Phase Reductions of Oscillators in Neuroscience, Electronics, and Physics}
  \label{sidebar:phase-reduction}

  \sdbarinitial{O}scillatory networks are omnipresent in both natural and engineered systems. However, 
  unlike the simplified phase dynamics~\eqref{eq:kuramoto} and~\eqref{eq:stochastic-kuramoto},
  models of such systems are generally expressed in terms of variables relevant to their 
  physical operation e.g. voltages and currents in analog circuits, 
  firing rates or ion channel activations in neurodynamics. This raises the 
  question: \textbf{when do the phase dynamics of real-world oscillators correspond to the 
  models~\eqref{eq:kuramoto} or~\eqref{eq:stochastic-kuramoto} used for analog computation?}
  The answer can be obtained using a technique called 
  \emph{phase reduction}~\cite{BM-DW-TM-JM:19, BP-AD:19}, which rigorously reduces
  state-space models to a representation involving only the phase of the oscillation.  

  \section{Phase Reduction of a Single Oscillator}
  We begin by considering a single oscillator, with an input, whose state encodes physically 
  relevant quantities
  \begin{sequation}
    \label{seq:single-dynamics}
    \dot{x} = f(x) + \epsilon u.
  \end{sequation}
  We assume that when $u \equiv 0$,
  \begin{enumerate}
    \item There exists an asymptotically stable 
    $T$-period limit cycle with basin of attraction $\Bc$, denoted $x^*(t)$. 
    \item There exists an \emph{asymptotic phase}, i.e. 
    a map $\theta:\Bc \to [0, 2\pi]$ such that for all initial 
    conditions $x_0 \in \Bc$, the solution $x(t)$ satisfies
    \[ \lim_{t \to \infty} \norm{x(t) - x^*\Big(t + \frac{T}{2\pi}\theta(x_0)\Big)} = 0  \]
  \end{enumerate}

  The asymptotic phase can be interpreted as the phase difference between the state 
  and a fixed reference point as it asymptotically converges to the 
  limit cycle. We are interested in characterizing how a weak input affects 
  the phase of the limit cycle oscillation. 
  
  If $x$
  is close to the limit cycle, 
  we can approximate~$x$ with the point exactly on the limit 
  cycle having the same phase, 
  $x \approx x^*\Big(\frac{T}{2\pi}\theta\big(x\big) \Big)$.
  Then $\theta(x(t))$ evolves according to
  \[ 
    \begin{aligned}
      \ddfrac{}{t}\theta(x) 
      &=  \nabla \theta(x)^\top \big(\,f(x) + \epsilon u\big) \\
      &\approx \omega + \epsilon \nabla \theta\bigg(x^*\Big(\frac{T}{2\pi}\theta(x)\Big)\bigg)^\top u.
    \end{aligned} 
  \]  
  where in the previous expression we used the fact that 
  $\nabla \theta (x^*(t))^\top f(x^*(t)) = \omega$, since along 
  the limit cycle the phase evolves
  at a constant rate, i.e. $\ddfrac{}{t}\theta(x^*(t)) = \nabla \theta (x^*(t))^\top f(x^*(t)) = \omega$. 

  %
  %
  Let $Z(\cdot) = \nabla \theta(x^*(\frac{T}{2\pi}\cdot))$ be 
  the \emph{infinitesimal phase response curve} (iPRC). 
  By substituting~$Z(\theta)$ into the previous equation, we see that 
  the phase evolution can be modeled with an autonomous dynamical 
  system 
  \begin{sequation}
    \label{seq:absolute-phase-dynamics}
    \dot{\theta} = \omega + \epsilon Z(\theta)^\top u.
  \end{sequation}

  \sdbarfig{
    \centering
    \includegraphics[width=\linewidth]{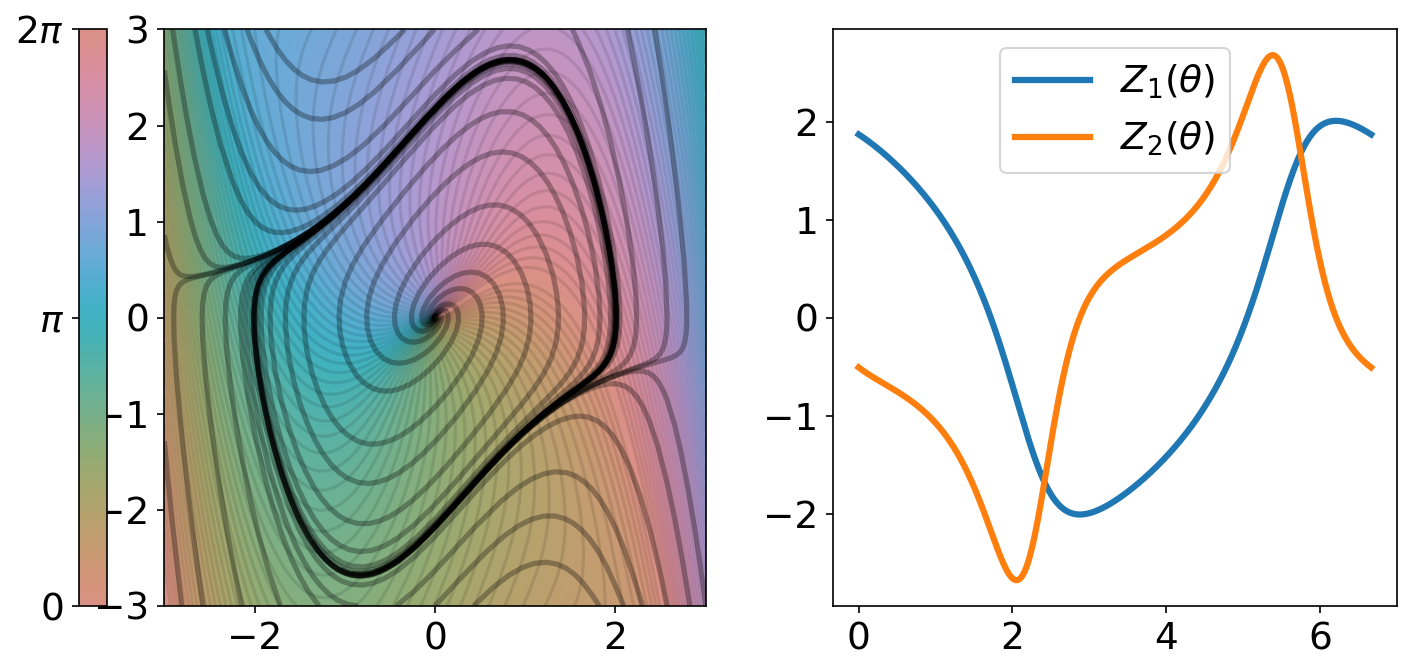}}{
      Asymptotic phase (left) and infinitesimal phase response curve (right) of the 
      Van der Pol oscillator $\dot{x}_1=x_2$, $\dot{x}_2 = \mu (1 - x_1)^2x_2 - x_1$\label{sfig:asymptotic-phase}  
    }

  In experimental settings, the iPRC may be obtained by
  subjecting the system to a small external perturbation and empirically 
  measuring the phase shift of the response~\cite{EMI:07,EB-JM-PH:04} (this 
  is referred to as the \emph{direct method}). 
  If a mathematical model 
  of the form~\eqref{seq:single-dynamics} is available, it is also possible to
  numerically calculate it using the \emph{adjoint method} \cite{BM-DW-TM-JM:19}, 
  which consists of solving the adjoint equation
  \begin{sequation}
    \label{seq:adjoint}
    \ddfrac{Z(\theta)}{\theta} = -\left.\pfrac{f(x)}{x}^\top\right|_{x = x^*(\frac{T}{2\pi}\theta)} Z(\theta) 
  \end{sequation}
  subject to the conditions that $f(x^*(0))^\top Z(0) = \omega$ and $Z$ 
  is $2\pi$-periodic. 
  Because the adjoint system is unstable when $x^*(t)$ is stable, 
  it must be integrated backward in time. 

  When $u \equiv 0$, the phase evolves at a constant rate $\theta(t) = \omega t$. 
  We define the \emph{phase deviation} as the difference between the phase 
  under the forced dynamics with respect to the unforced reference:
  $\phi = \theta - \omega t$. Then 
  \begin{sequation}
    \label{seq:phase-deviation-dynamics}
    \dot{\phi} = \epsilon Z(\phi + \omega t)^\top u
  \end{sequation}
  Suppose that $u$ is $T$-periodic. When $\epsilon$ is small, 
  the timescale over which the phase deviation
  changes is slow relative to the oscillatory frequency of $u$. Thus, 
  \eqref{seq:phase-deviation-dynamics} may be simplified by
  averaging the high-frequency oscillations
  and changing timescales $t \mapsto \frac{1}{\epsilon} t$ ~\cite[Section 9.3]{FCH-EMI:12}, \cite{JAS-FV-JM:07},
  \begin{sequation}
    \label{seq:average-phase-deviation}
    \dot{\phi} \approx \Omega(\phi) \qquad \Omega(\phi) := \frac{1}{T}\int_{0}^{T}Z(\phi + \omega t)^\top u(t)dt.
  \end{sequation}
  Note that equations similar to~\eqref{seq:phase-deviation-dynamics}
  and~\eqref{seq:average-phase-deviation} can 
  also be derived using Adler's equation which models injection locking 
  in analog circuits~\cite{RA:46, PB-JR:09-aspdac}. 
\end{sidebarpart}

\begin{sidebarpart}{\continuesidebar}
  \section{Phase Reduction of Oscillator Networks}
  We now extend this analysis to networks of limit cycle oscillators. 
  Suppose that we have a network with $N$ nodes whose adjacency matrix is 
  given by $J \in \real^{N \times N}$. We assume that 
  dynamics are \emph{weakly connected}, meaning they have the form
  \begin{sequation}
    \label{seq:weakly-connected-network}
      \dot{x}_i = f(x_i) + \epsilon \left(v_i + \sum_{j=1}^{N}J_{ij}g(x_i, x_j)\right) 
  \end{sequation}
  The system may be interpreted as $N$ copies 
  of~\eqref{seq:single-dynamics}, where the input to the $i$th node, 
  $u_i = v_i + \sum_{j=1}^{N}J_{ij}g(x_i, x_j)$, consists of an input applied directly to the node, 
  as well as its interactions with its neighbors. 

  Repeating the steps of the derivation in the single oscillator case, and letting 
  $\Gamma$ be the average of the nodes' interactions,
  \[ \Gamma(\Delta\phi) = \frac{1}{T}\int_{0}^{T}Z(\omega t)^\top g\Big(x^*(t), x^*\big(t + \frac{T}{2\pi}\Delta\phi\big)\Big)  \]
  we obtain 
  \begin{sequation}
    \label{seq:general-coupled-oscillator-network}
    \dot{\phi}_i = \Omega(\phi_i) + \sum_{j=1}^{N}J_{ij}\Gamma(\phi_j - \phi_i),
  \end{sequation}
  which is a generalized form of~\eqref{eq:kuramoto}. Extensions of phase reduction 
  techniques to stochastic oscillators~\cite{JT-HN-GBE:09, ZA-PH-VS:19, ZA-VS:22} may also be used to obtain models 
  of the form~\eqref{eq:stochastic-kuramoto}
  \begin{sequation}
    \label{seq:general-stochastic-coupled-oscillator-network}
    \dot{\phi}_i = \Omega(\phi_i) + \sum_{j=1}^{N}J_{ij}\Gamma(\phi_j - \phi_i) + dw_i,
  \end{sequation}
  
  Models of the form~\eqref{seq:general-coupled-oscillator-network} 
  and~\eqref{seq:general-stochastic-coupled-oscillator-network}
  abstract away the physics from real-world instances of coupled oscillator networks,
  and convert them into forms more suitable for reasoning about their computational 
  capabilities.
  \putbib[refs]
  \pagebreak
\end{sidebarpart}
\end{bibunit}

\section{Combinatorial Optimization with Coupled Oscillator Networks}
We now move on to the first computing application of coupled oscillator 
networks, solving combinatorial optimization problems. Examples include the 
MAX-CUT problem, the traveling salesman problem, and 
the Boolean satisfiability problem, and are 
of importance both in theoretical computer science, as well as practical 
applications such as operations research, VLSI circuit design, and 
artificial intelligence~\cite{FB-MG-MJ-GR:88, CA-SB-PK-SNM:10-eccv, JW-TJ-SFC:13}.
Large instances 
of such problems are generally intractable on digital computers due 
to the fact that they are NP-complete or NP-hard. 

However, as we show here, coupled oscillator 
networks are capable of solving these problems efficiently by 
exploiting the fact that the solutions to many NP problems can be encoded 
in the ground states of a 
corresponding Ising Hamiltonian \cite{AL:14} 
(see, e.g.,
Sidebar: \sidebarref{sidebar:ising-np}). 
Physical devices that minimize the Hamiltonian, called Ising machines, 
can be used as hardware solvers of combinatorial optimization problems~\cite{NM-PLM-TB:22}.
Ising machines based on oscillator networks have been proposed to solve a number of 
NP-hard problems including the graph coloring
problem~\cite{CWW:98, JW-LJ-RL-WC:11, AP-NS-MJ-SD-AR:17}, 
Boolean satisfiability problem~\cite{MB-ZL-NS:23}, and the 
graph maximum cut problem~\cite{SD-AK-AA-HP-DS-ZT-AR-SD:21}.

\begin{sidebarfixed}[p]{Ising Models of NP Problems}
  \label{sidebar:ising-np}
  \sdbarinitial{T}he Ising model provides a framework to reason about many problems in computing, 
  since these problems are often equivalent to computing the ground states of 
  a certain Ising Hamiltonian. We show here the sense in which minimizing an Ising Hamiltonian 
  can serve as a proxy for ``difficult'' problems in computer science. 

  Consider 
  the following combinatorial optimization problem involving an Ising Hamiltonian~\eqref{eq:ising-hamiltonian},
  \begin{sequation}
    \label{seq:ising-optimization}
    \min_{s \in \{-1, 1\}^N} H(s).
  \end{sequation}
  It is useful to distinguish between two types of problems related to minimizing the Ising Hamiltonian~\cite{MM-AM:09}:
  \begin{enumerate}
    \item \textbf{Ising Decision Problem}: Given a value $H_0 \in \real$, establish whether there exists 
    a spin configuration $s \in \{-1, 1\}^N$ such that $H(s) < H_0$. 
    \item \textbf{Ising Optimization Problem}: Identify a spin configuration $s^* \in \{-1, 1\}^N$ solving~\eqref{seq:ising-optimization}
  \end{enumerate}
  Given a ``difficult'' problem $\Qc$, the formulation of the Ising problem one uses as a proxy depends on the 
  form of $\Qc$: if $\Qc$ is a decision problem, i.e. it can be posed as a question requiring a ``yes'' or ``no'' answer, 
  the Ising Decision problem is the natural proxy, whereas if $\Qc$ is naturally formulated as a search problem, 
  one may instead use the optimization formulation. In either case, an Ising machine solves 
  the Ising Optimization problem, though its solution can be used to solve the Ising Decision Problem.

  Computing problems are sorted into \emph{complexity classes} based on how the effort 
  required to solve them scales asymptotically as $N \to \infty$. A decision problem belongs 
  to \emph{class P} if it can be solved in polynomial time, i.e. 
  the number of steps needed to solve it is polynomial in $N$. It
  belongs to \emph{class NP} if there exists a certificate one can use to validate
  the solution in polynomial time. Any problem in class P is also in class NP, but whether the reverse 
  inclusion holds is still an open problem in theoretical computer science. 

  \sdbarfig{
    \drawPneqNP\hspace{1em}\drawPeqNP
  }{Relationship between computational complexity classes in the cases 
  where $\text{P}\neq\text{NP}$ (left) and $\text{P}=\text{NP}$ (right).\label{sfig:P-vs-NP}}    

  Following the convention in~\cite{AL:14}, we delineate two additional complexity classes that correspond to the most 
  difficult computing problems, and the formulation of the Ising problem that serves as a proxy:
  \begin{itemize}
    \item \textbf{NP-complete:} A problem $\Qc$ is called NP-complete if $\Qc$ is in class 
    NP and any other NP problem can be reduced to $\Qc$ in polynomial time. Thus if one can efficiently solve 
    $\Qc$, any other NP problem can also be solved quickly. \emph{NP-complete problems are
    equivalent to the Ising Decision Problem given a corresponding Hamiltonian}, such as 
    the Boolean satisfiability problem. 
    
    \item \textbf{NP-hard:} A problem $\Qc$ is called NP-hard if any other NP problem can 
    be reduced to $\Qc$ in polynomial time, though $\Qc$ does not necessarily need to belong to 
    class NP. \emph{Many NP-hard problems are equivalent to the Ising Optimization Problem given a 
    corresponding Hamiltonian}, such as the graph partitioning problem. 
  \end{itemize}
  The relationship between these complexity classes is illustrated in Figure~\ref{sfig:P-vs-NP}. We 
  refer the reader to~\cite{AL:14} for a comprehensive survey of Ising formulations of common 
  NP-complete and NP-hard problems.
\end{sidebarfixed}

Formally, we are interested in solving the problem of minimizing an Ising 
Hamiltonian
\begin{equation}
  \label{eq:minimize-ising}
  \min_{s \in \{-1, 1\}^N} H(s) := -\frac{1}{2}\sum_{i=1}^{N}\sum_{j=1}^{N}J_{ij}s_is_j
\end{equation}
The system we use to solve~\eqref{eq:minimize-ising} is a Kuramoto-like system of coupled 
phase oscillators introduced in~\cite{TW-JR:19}, called an \emph{Oscillator Ising Machine} (OIM), 
whose dynamics are 
\begin{equation}
  \label{eq:forced-kuramoto-2}
    \dot{\phi}_i = \sum_{j=1}^{N}J_{ij}\sin(\phi_j - \phi_i) - \mu_i \sin(2\phi_i)  \qquad i=1, \dots, N
\end{equation}
which is the gradient flow of 
the energy function
\begin{equation}
  \label{eq:forced-kuramoto-energy-2}
  E_\mu(\phi) = -\frac{1}{2}\sum_{i=1}^N\sum_{j=1}^N J_{ij}\cos(\phi_j - \phi_i) + \sum_{i=1}^{N}\mu_i \sin^2(\phi_i).
\end{equation}
This system converges to equilibria corresponding to spin configurations 
of the Ising model through 
the map $\phi^*:\{-1, 1\}^N \to \Tb^N$, where 
\begin{equation}
  \label{eq:phase-configuration-2}
  \phi_i^*(s) = \begin{cases}
    0 \qquad s_i = \,+\>\!1 \\ \pi \qquad s_i = -1.
  \end{cases}
\end{equation}
Note that \eqref{eq:forced-kuramoto-2} may possess equilibria 
that do not have the form \eqref{eq:phase-configuration-2}, though in general 
such equilibria are structurally unstable~\cite{YC-MBK-NS-ZL:24}.

The forcing term $\mu_i\sin(2\phi_i)$ is called 
\emph{second harmonic injenction locking}
since in circuit realizations it is generated by an exogenous periodic input whose frequency 
is twice the natural frequency of the oscillator at node~$i$. It serves 
to penalize phase values~$\phi_i \not\in \{0, \pi \}$ that do not correspond to valid 
spin configurations. 
The regularization parameters~$\mu_i \geq 0$ adjust the degree of penalization. 
While many analyses of OIMs assume that $\mu$ is homogeneous across 
all nodes, recent work~\cite{AA-ANM-FP-AM:25-cdc} has 
established that, in many cases, 
heterogeneity improves convergence to low energy spin configurations. 

By definition, $E_\mu(\phi^*(s)) = H(s)$ for all spin configurations~$s$.
Thus minimizing $E_\mu$ may serve as a proxy for minimizing the Ising Hamiltonian, 
and in practice one observes good correspondence between solutions computed by OIMs
and ground states of the Ising model.
Empirical benchmarks show that OIMs 
realized as conventional CMOS circuits achieve superior performance and scalability to 
classical algorithms, such as simulated annealing, as well as non-oscillatory 
Ising machines such as the D-WAVE quantum Ising machine and laser-based coherent Ising 
machines~\cite{TW-LW-PN-JR:21}. However theoretical guarantees on the performance 
of OIMs are still lacking. This is because the relationship between solutions
of~\eqref{eq:minimize-ising} and the stability of equilibria 
of~\eqref{eq:forced-kuramoto-2} is highly complex. In particular, 
\emph{$s^*$ being a global optimizer of the Ising Hamiltonian is neither 
necessary nor sufficient 
for stability of the corresponding equilibrium $\phi^*(s^*)$}, 
as we show in the following example. 
\begin{csm_example}
\label{ex:minimum-is-not-stable}
Consider the frustrated network in Figure~\ref{fig:frustration}. 
We have the following:
\begin{itemize}
  \item If $\mu_i = 0.1$ for all $i$, then~\eqref{eq:forced-kuramoto-2} does not have any stable 
  equilibria corresponding to a valid spin configuration, and in particular 
  $\phi^*(s^*)$ is unstable for every global minimum $s^*$. 
  \item  If $\mu_i = 10$ for all $i$, then $\phi^*(s)$ is asymptotically stable for all 
  $s \in \{-1, 1 \}^N$, implying that the equilibrium corresponding to a spin configuration is asymptotically stable regardless of 
  whether it is a minimizer of~\eqref{eq:minimize-ising} or not.
  \item There exists no $\mu$ such that the only stable equilibria of~\eqref{eq:forced-kuramoto-2}
  correspond to minimizers of~\eqref{eq:minimize-ising}. \oprocend
\end{itemize}
\end{csm_example}
The previous example illustrates the difficulties related to stability analysis 
of~\eqref{eq:forced-kuramoto-2}, and in particular, the impact 
of the choice of $\mu$: if $\mu$
is too small, the OIM fails to converge to an equilibrium 
corresponding to a valid spin configuration of the 
Ising Hamiltonian, while if $\mu$ is too large, the OIM
also stabilizes spin configurations\footnote{We sometimes abuse terminology and use the 
term ``spin configuration'' to refer both to $s \in \{-1, 1\}^N$ 
and the corresponding equilibrium,
$\phi^*(s)$, of \eqref{eq:forced-kuramoto-2}.} that are suboptimal.
As we show later, this challenge is intrinsically related 
to the presence of frustration in the network, which together 
with the combinatorial explosion in the number of equilibria of 
large networks (in a network with $N$ nodes, there are $2^N$ spin configurations), 
means characterizing precisely the relationship between the stability 
and optimality of a given spin configuration is a nontrivial problem. 

In the subsequent section we introduce a framework to analyze the 
stability and optimality of spin 
configurations
using methods from the 
theory of signed graphs. Given a fixed network, our approach associates
each spin configuration, $s$, to a signed graph,~$\Gc(s)$, 
where both stability and optimality of~$\phi^*(s)$ are encoded
in the eigenvalues of its corresponding Laplacian matrix. Thus, 
by analyzing the spectra of these Laplacians, we can characterize the sense 
in which low energy spin configurations are ``more stable'', as well 
as the effect of the regularization parameters. 
For frustration-free networks we obtain
exact stability results, and for frustrated networks we establish 
this relationship probabilistically. This formalism also allows us to 
investigate the role of heterogeneity on stability properties, and in particular, 
show that heterogeneous networks preferentially stabilize low 
energy spin configurations. 

\subsection{Signed Graphs of the Ising Model}
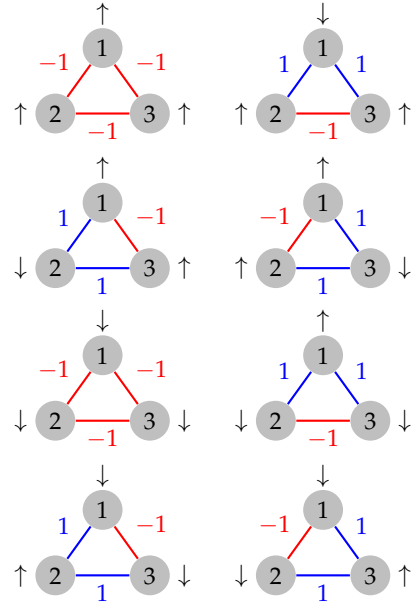
\begin{figure}
  \begin{figframe}
  \begin{tikzpicture}[node distance=13pt]
    \tikzstyle{spinsite}=[circle,fill=black!25,minimum size=15pt,inner sep=0pt, node distance=120pt]

    \node (n1) at (0, 0) [spinsite] {$1$};
    \node (n2) at (1.25*0.5, -0.866*1.0) [spinsite] {$3$};
    \node (n3) at (-1.25*0.5, -0.866*1.0) [spinsite] {$2$};

    \draw[thick, color=red] (n1) -- (n2) node[midway, above right] {$-1$};
    \draw[thick, color=red] (n2) -- (n3) node[midway, below] {$-1$};
    \draw[thick, color=red] (n3) -- (n1) node[midway, above left] {$-1$};

    \node[above of=n1] {$\uparrow$};
    \node[right of=n2] {$\uparrow$};
    \node[left of=n3] {$\uparrow$};
  \end{tikzpicture}
  \hspace{1.0ex}
  \begin{tikzpicture}[node distance=13pt]
    \tikzstyle{spinsite}=[circle,fill=black!25,minimum size=15pt,inner sep=0pt, node distance=120pt]

    \node (n1) at (0, 0) [spinsite] {$1$};
    \node (n2) at (1.25*0.5, -0.866*1.0) [spinsite] {$3$};
    \node (n3) at (-1.25*0.5, -0.866*1.0) [spinsite] {$2$};

    \draw[thick, color=blue] (n1) -- (n2) node[midway, above right] {$1$};
    \draw[thick, color=red] (n2) -- (n3) node[midway, below] {$-1$};
    \draw[thick, color=blue] (n3) -- (n1) node[midway, above left] {$1$};

    \node[above of=n1] {$\downarrow$};
    \node[right of=n2] {$\uparrow$};
    \node[left of=n3] {$\uparrow$};
  \end{tikzpicture}
  \hspace{1.0ex}
  \begin{tikzpicture}[node distance=13pt]
    \tikzstyle{spinsite}=[circle,fill=black!25,minimum size=15pt,inner sep=0pt, node distance=120pt]

    \node (n1) at (0, 0) [spinsite] {$1$};
    \node (n2) at (1.25*0.5, -0.866*1.0) [spinsite] {$3$};
    \node (n3) at (-1.25*0.5, -0.866*1.0) [spinsite] {$2$};

    \draw[thick, color=red] (n1) -- (n2) node[midway, above right] {$-1$};
    \draw[thick, color=blue] (n2) -- (n3) node[midway, below] {$1$};
    \draw[thick, color=blue] (n3) -- (n1) node[midway, above left] {$1$};

    \node[above of=n1] {$\uparrow$};
    \node[right of=n2] {$\uparrow$};
    \node[left of=n3] {$\downarrow$};
  \end{tikzpicture}
  \hspace{1.0ex}
  \begin{tikzpicture}[node distance=13pt]
    \tikzstyle{spinsite}=[circle,fill=black!25,minimum size=15pt,inner sep=0pt, node distance=120pt]

    \node (n1) at (0, 0) [spinsite] {$1$};
    \node (n2) at (1.25*0.5, -0.866*1.0) [spinsite] {$3$};
    \node (n3) at (-1.25*0.5, -0.866*1.0) [spinsite] {$2$};

    \draw[thick, color=blue] (n1) -- (n2) node[midway, above right] {$1$};
    \draw[thick, color=blue] (n2) -- (n3) node[midway, below] {$1$};
    \draw[thick, color=red] (n3) -- (n1) node[midway, above left] {$-1$};

    \node[above of=n1] {$\uparrow$};
    \node[right of=n2] {$\downarrow$};
    \node[left of=n3] {$\uparrow$};
  \end{tikzpicture}


  \begin{tikzpicture}[node distance=13pt]
    \tikzstyle{spinsite}=[circle,fill=black!25,minimum size=15pt,inner sep=0pt, node distance=120pt]

    \node (n1) at (0, 0) [spinsite] {$1$};
    \node (n2) at (1.25*0.5, -0.866*1.0) [spinsite] {$3$};
    \node (n3) at (-1.25*0.5, -0.866*1.0) [spinsite] {$2$};

    \draw[thick, color=red] (n1) -- (n2) node[midway, above right] {$-1$};
    \draw[thick, color=red] (n2) -- (n3) node[midway, below] {$-1$};
    \draw[thick, color=red] (n3) -- (n1) node[midway, above left] {$-1$};

    \node[above of=n1] {$\downarrow$};
    \node[right of=n2] {$\downarrow$};
    \node[left of=n3] {$\downarrow$};
  \end{tikzpicture}
  \hspace{1.0ex}
  \begin{tikzpicture}[node distance=13pt]
    \tikzstyle{spinsite}=[circle,fill=black!25,minimum size=15pt,inner sep=0pt, node distance=120pt]

    \node (n1) at (0, 0) [spinsite] {$1$};
    \node (n2) at (1.25*0.5, -0.866*1.0) [spinsite] {$3$};
    \node (n3) at (-1.25*0.5, -0.866*1.0) [spinsite] {$2$};

    \draw[thick, color=blue] (n1) -- (n2) node[midway, above right] {$1$};
    \draw[thick, color=red] (n2) -- (n3) node[midway, below] {$-1$};
    \draw[thick, color=blue] (n3) -- (n1) node[midway, above left] {$1$};

    \node[above of=n1] {$\uparrow$};
    \node[right of=n2] {$\downarrow$};
    \node[left of=n3] {$\downarrow$};
  \end{tikzpicture}
  \hspace{1.0ex}
  \begin{tikzpicture}[node distance=13pt]
    \tikzstyle{spinsite}=[circle,fill=black!25,minimum size=15pt,inner sep=0pt, node distance=120pt]

    \node (n1) at (0, 0) [spinsite] {$1$};
    \node (n2) at (1.25*0.5, -0.866*1.0) [spinsite] {$3$};
    \node (n3) at (-1.25*0.5, -0.866*1.0) [spinsite] {$2$};

    \draw[thick, color=red] (n1) -- (n2) node[midway, above right] {$-1$};
    \draw[thick, color=blue] (n2) -- (n3) node[midway, below] {$1$};
    \draw[thick, color=blue] (n3) -- (n1) node[midway, above left] {$1$};

    \node[above of=n1] {$\downarrow$};
    \node[right of=n2] {$\downarrow$};
    \node[left of=n3] {$\uparrow$};
  \end{tikzpicture}
  \hspace{1.0ex}
  \begin{tikzpicture}[node distance=13pt]
    \tikzstyle{spinsite}=[circle,fill=black!25,minimum size=15pt,inner sep=0pt, node distance=120pt]

    \node (n1) at (0, 0) [spinsite] {$1$};
    \node (n2) at (1.25*0.5, -0.866*1.0) [spinsite] {$3$};
    \node (n3) at (-1.25*0.5, -0.866*1.0) [spinsite] {$2$};

    \draw[thick, color=blue] (n1) -- (n2) node[midway, above right] {$1$};
    \draw[thick, color=blue] (n2) -- (n3) node[midway, below] {$1$};
    \draw[thick, color=red] (n3) -- (n1) node[midway, above left] {$-1$};

    \node[above of=n1] {$\downarrow$};
    \node[right of=n2] {$\uparrow$};
    \node[left of=n3] {$\downarrow$};
  \end{tikzpicture}
  \end{figframe}
  \caption{Signed graphs corresponding to spin configurations of the network $J=[0, -1, -1; -1, 0, -1; -1, -1, 0]$. 
  Because the network is frustrated, no configuration results in a signed graph with only positive edges.}
  \label{fig:signed-graphs}
\end{figure}

Here we introduce a framework to analyze the stability 
of~\eqref{eq:forced-kuramoto-2} using methods from the 
theory of signed graphs, originally proposed 
in~\cite{AA-ANM-FP-AM:25-cdc}. Let $J$ be a fixed adjacency matrix 
encoding interactions between nodes. For all $s \in \{-1, 1 \}^N$,
let $\Gc(s)$ be the signed graph with adjacency matrix 
\begin{equation}
  \label{eq:signed-adjacency}
  A_{ij}(s) = J_{ij}s_is_j
\end{equation}
In Figure \ref{fig:signed-graphs}, we demonstrate the signed graphs corresponding to 
the frustrated network from Example~\ref{ex:minimum-is-not-stable}. 
By construction, the sign of an edge $(i, j)$ obeys the following rule
\[
  \begin{aligned}
    \operatorname{sgn}(A_{ij}(s)) = 
    \begin{cases}
      \phantom{-}1 \;\; &\text{local energy $(i, j)$ is minimized} \\
      -1 \;\; &\text{local energy $(i, j)$ is maximized}
    \end{cases}
  \end{aligned}
\]
Properties of the Ising model can be recast within this signed graph theoretic 
framework. A network is \emph{frustrated} if for all spin configurations 
$s \in \{-1, 1 \}^N$ the corresponding signed graph $\Gc(s)$ 
contains at least one negative edge, otherwise the 
network is \emph{frustration-free}. 

The Laplacian of a signed graph is the matrix $L(s) = D(s) - A(s)$, where 
$D(s)$ is the diagonal matrix consisting of row sums of $A(s)$. We 
have 
\begin{equation}
  \label{eq:signed-laplacian}
  \begin{aligned}
    L_{ij}(s) = 
    \begin{cases}
      \sum_{k=1}^{N}J_{ik}s_is_k \qquad &i=j \\ 
      -J_{ij}s_is_j \qquad &i \neq j
    \end{cases}
  \end{aligned}
\end{equation}
Both optimality and stability 
of a spin configuration $s$ are related to the spectral 
properties of the matrix~$L(s)$. The Ising Hamiltonian satisfies
\[ H(s) = -\frac{1}{2}\operatorname{trace}(L(s)) = -\frac{1}{2}\sum_{i=1}^{N}\lambda_i(L(s)) \]
where $\lambda_1(L(s)) \leq \lambda_2(L(s)) \leq \cdots \leq \lambda_N(L(s))$
are the eigenvalues of $L(s)$. 
Since \eqref{eq:forced-kuramoto-2} is the 
gradient system of the energy function~$E_\mu$, 
stability can be determined from the eigenvalues of 
the Hessian of $E_\mu$. The spin configuration is 
asymptotically stable when $\nabla^2 E_\mu(\phi^*(s)) \succeq 0$, 
or equivalently, when $\lambda_1( \nabla^2 E_\mu(\phi^*(s))) > 0$ where 
$\lambda_1$ denotes the smallest eigenvalue. 
By direct computation, 
\begin{equation}
  \label{eq:hessian}
  \begin{aligned}
    &(\nabla^2 E_\mu(\phi))_{ij} = \\
    &\qquad\begin{cases}
      \sum_{k=1}^{N}J_{ik}\cos(\phi_k - \phi_i)+\,2\mu_i\cos(2\phi_i) \;\; &i=j\\
      -J_{ij}\cos(\phi_j - \phi_i) \qquad &i \neq j,
    \end{cases}
  \end{aligned}
\end{equation}
and evaluating \eqref{eq:hessian} at the equilibrium $\phi^*(s)$
yields
\[  \nabla^2 E_\mu(\phi^*(s)) = L(s) + 2\operatorname{diag}(\mu_1, \dots, \mu_N). \]

Thus the connection between the stability of a 
spin configuration and its energy value corresponds
to the relationship between the sum of the 
eigenvalues of~$L(s)$ and the minimum eigenvalue 
of the perturbed matrix $L(s) + 2\diag(\mu)$. In the 
following we characterize this relationship exactly in the
frustration-free case, and probabilistically in the frustrated 
case. 

\subsection{Stability of Frustration-free Networks}
We now analyze the stability of the system~\eqref{eq:forced-kuramoto-2}
using the signed graph theoretic framework introduced in the previous 
section, in the special case where the network $J$ is frustration-free, 
utilizing the notion of \emph{structural balancing} of signed graphs, 
which plays an important role in dynamics on signed networks~\cite{CA:13,GS-CA-JSB:19}. 

Let $s \in \{-1, 1\}^N$ be a spin configuration and $\Gc(s)$ 
its corresponding signed graph. We say that $\Gc(s)$ is \emph{structurally balanced}
if there exists a partition of the set of nodes $\Vc = \{1, 2, \dots, N \}$ 
into $\Vc_1 \subset \Vc$ and $\Vc_2 \subset \Vc$ such that edges within $\Vc_1$ or $\Vc_2$ are 
positive, and edges between $\Vc_1$ and $\Vc_2$ are negative. The corresponding partition 
is called \emph{balanced}. 
We have the following result.
\begin{lemma}
  \label{lem:sign-balanced-frustration-free}
  Let $s \in \{-1, 1 \}^N$ be a spin configuration. Then $J$ 
  is frustration-free if and only if $\Gc(s)$ is structurally balanced.
\end{lemma}

\begin{figure}
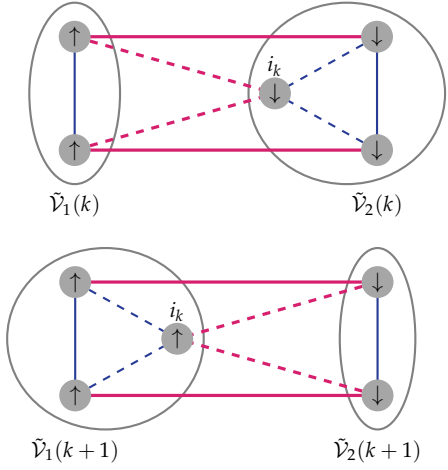

  \begin{figframe}
    \center
    \partitionswap{1} \\
    \partitionswap{2}
  \end{figframe}
  \caption{Spin flipping procedure described in proof of Lemma~\ref{lem:sign-balanced-frustration-free}. 
  Positive edges are blue and 
  negative edges are red. Edges incident to~$i_k$ are depicted by dashed lines. 
  (Top) Structurally balanced graph partitioned into sets~$\tilde{\Vc}_1(k)$ and~$\tilde{\Vc}_2(k)$. 
  (Bottom) Flipping the spin at~$i_{k}$ also flips the signs of all of its edges. To maintain 
  balancing of the partition, $i_{k}$ is transferred from $\tilde{\Vc}_2$ to $\tilde{\Vc}_1$ 
  on the $(k + 1)$th iteration.}
  \label{fig:cut-swap}
\end{figure}

\begin{proof}
  We begin with the forward direction. If~$J$ is frustration-free, there exists 
  a spin configuration~$s^* \in \{-1, 1 \}^N$ such that~$J_{ij}s_i^*s_j^* > 0$
  for all edges $(i, j)$. Letting $\Vc_1 = \{i \mid s_i = s^*_i \}$ 
  and $\Vc_2 = \{i \mid s_i \neq s^*_i \}$, it follows immediately 
  that this partition is \emph{balanced}. 

  To prove the reverse direction, suppose $\Gc(s)$ is structurally balanced, 
  and let $\Vc_1, \Vc_2$ be the corresponding partition of the node set. If 
  either $\Vc_1 = \emptyset$ or $\Vc_2 = \emptyset$, then all edges within 
  the signed graph are positive, and the result follows immediately. 
  
  For the case where both sets are nonempty, we introduce an iterative procedure to 
  construct a spin configuration 
  where each interaction energy is simultaneously minimized. Let 
  $\tilde{s}(0) = s$, $\tilde{\Vc}_1(0) = \Vc_1$, and $\tilde{\Vc}_2(0) = \Vc_2$. 
  For each iteration $k \in \intpos$, choose a node~$i_k \in \tilde{\Vc}_2(k)$, and let
  \[
    \begin{aligned}
      \tilde{s}_j(k + 1) &= 
        \begin{cases}
          -s_j(k)  \qquad j \neq i_k \\ 
          \phantom{-}s_j(k) \qquad j = i_k.
        \end{cases} \\
      \tilde{\Vc}_1(k + 1) &= \tilde{\Vc}_1(k) \cup \{i_k \} \\
      \tilde{\Vc}_2(k + 1) &= \tilde{\Vc}_2(k) \setminus \{i_k \}
    \end{aligned}
  \]
  The idea is that at each iteration, a new spin configuration is obtained by flipping 
  the spin of node $i_k \in \tilde{\Vc}_2(k)$. Because flipping the spin at node $i_k$ also flips the signs of all the 
  edges incident to $i_k$, we transfer $i_k$ from $\tilde{\Vc}_2$ to the set $\tilde{\Vc}_1$
  in order to maintain balancing. We illustrate this procedure 
  in Figure~\ref{fig:cut-swap}. Since $\tilde{\Vc}_2(k) = \emptyset$ at iteration~$k=|\tilde{\Vc}_2(0)|$,  
  all edges of~$\Gc(\tilde{s}(k))$ are positive and $J$ is frustration-free. 
\end{proof}

The previous result establishes equivalence between structural balance 
of the signed graphs $\Gc(s)$ and $J$ being frustration-free. In the following result, 
we leverage this connection to analyze the stability of minimizers of 
the Ising Hamiltonian by applying results on the spectra of Laplacians 
corresponding to structurally balanced signed graphs. Specifically, we identify 
conditions on~$\mu$ such that~$\phi^*(s)$ is asymptotically 
stable if and only if~$\phi^*(s)$ is a global minimizer of \eqref{eq:minimize-ising}. 

\begin{theorem}
  Let $J$ be a frustration-free network. If $0 < \mu_i < \mu^*$ for all $i=1, \dots, N$, where $\mu^*$ 
  is given by~\eqref{eq:mu-upper-bound}, it follows that $\phi^*(s)$ is asymptotically stable if and only if 
  $s \in \{-1,  1\}^N$ is a global minimizer of \eqref{eq:minimize-ising}. 
\end{theorem}

\begin{proof}
  Let $s \in \{-1, 1\}^N$. If~$s$ is a global minimizer of \eqref{eq:minimize-ising}, 
  then~$s$ simultaneously minimizes all local interaction energies, so~$\Gc(s)$ has only positive edges. 
  Therefore $L(s)$ corresponds to the Laplacian of an unsigned graph, so $L(s) \succeq 0$ (see~\cite[Chapter 13]{CDG-GFR:01}
  or \cite[Chapter 6]{FB-LNS} for details on the spectral properties of the graph Laplacian).
  Thus for any $\mu \in \real^{N}$, we have $ \nabla^2 E_\mu(\phi^*(s)) = L(s) + 2\diag(\mu) \succ 0$ and 
  therefore~$\phi^*(s)$ is stable. 

  If $s$ is not a global minimizer, then by Lemma~\ref{lem:sign-balanced-frustration-free} it 
  follows that $\Gc(s)$ is structurally balanced and has at least one negative edge. 
  By \cite[Corollary IV.4]{DZ-MB:17}, $L(s)$ is indefinite, thus $\lambda_1(L(s)) < 0$. 
  Letting 
  \begin{equation}
    \label{eq:mu-upper-bound}
    \mu^* = \min_{s' \in \{-1, 1 \}^N}\left\{-\frac{\lambda_1(L(s'))}{2} \mid s' \text{ is suboptimal} \right\},
  \end{equation}
  it follows that 
  \[ \lambda_1( \nabla^2 E_\mu(\phi^*(s))) = \lambda_1(L(s) + 2\diag(\mu)) < 0\] 
  for any $\mu \in \real^N_{> 0}$ where $\mu_i < \mu^*$, and $s$ is unstable.
\end{proof}

\subsection{Stability of Frustrated Networks}
We now move on to the problem of analyzing the stability of \eqref{eq:forced-kuramoto-2}
when the network $J$ is frustrated. In this case, the signed graphs corresponding to spin configurations 
are not structurally balanced, so analytic results on the spectrum of the graph Laplacian 
are unavailable. We overcome this challenge by analyzing ensembles of random networks rather 
than specific instances, and characterizing the relationship between the energy and stability 
of spin configurations statistically. 

For simplicity, we focus on 
the special case where all edges are 
antiferromagnetic, though it is straightforward to extend 
our approach to the setting where ferromagnetic interactions are present as well. 
In our model, both the network interconnection~$J$ and the spin configuration~$s$ 
are sampled randomly. To examine the effect of heterogeneity 
in the regularization parameters, $\mu$ is chosen randomly as well. 
Let $N > 0$ be the number of nodes in the graph and $0 < p < 1$. Then
\begin{enumerate}
  \item A network $J$ is randomly constructed following 
  the Erd\H{o}s--R\'enyi model, where for $1 \leq i < j \leq N$,
  \[ 
    J_{ij} = 
      \begin{aligned}
        \begin{cases}
        -1 \qquad &\text{with probability $p$} \\
         \phantom{-}0 \qquad &\text{with probability $1 - p$}
        \end{cases} 
      \end{aligned}
  \]
  \item A spin configuration $s \in \{-1, 1\}^N$ is sampled uniformly at random. 
  \item The regularization parameters $\mu_1, \dots, \mu_N$ are sampled independently 
  from a distribution $\Mc$ supported on~$[0, \infty)$ with mean $\bar{\mu}$ and variance $\sigma^2$. 
\end{enumerate}
Let $\Gc(s)$ be the signed graph whose adjacency matrix 
is~\eqref{eq:signed-adjacency}. Then, the graph Laplacian~$L(s)$, 
and the Hessian~$\nabla^2 E_\mu(\phi^*(s))$ are
random matrices, and their eigenvalues can be treated as random variables. 

The distribution of eigenvalues of the Laplacian and the Hessian can 
be derived using methods from random matrix theory \cite{TT:12, MP-JPB:20}. 
Given a symmetric matrix~$A \in \real^{N \times N}$, 
define the \emph{empirical spectral density} (ESD) of A as
\[ \rho(\lambda; A) = \frac{1}{N}\sum_{i=1}^{N}\delta(\lambda - \lambda_i(A)) \]
In the case where $A(s)$ is the adjacency matrix of $\Gc(s)$, it follows by 
classical results in random matrix theory\footnote{Specifically, $J/\sqrt{pN}$ is a \emph{Wigner ensemble}, 
meaning its entries are i.i.d. with zero mean and unit variance, and
by \cite[Theorem 2.4.2]{TT:12}, $\rho(\lambda; J) \to \rho_{\rm sc}$. Since 
$A(s) = S^{-1}JS$ where $S = \diag(s_1, \dots, s_N)$, the ESD of $A(s)$ converges to the same limit.}
that $\rho(\lambda; A(s) / \sqrt{pN})$ 
converges almost surely to 
the \emph{semi-circle law} $\rho_{\rm sc}(\lambda) = \frac{1}{2\pi}\sqrt{4 - \lambda^2}$ 
as $N \to \infty$. 
Recent work has also characterized the limit of the ESD 
of Laplacian matrices of random graphs \cite{WB-AD-TJ:06, XD-TJ:10, TJ:12}, 
however a key assumption in these works is that the entries of the adjacency matrix are
independent and identically distributed, which is violated for ensembles corresponding to 
signed graphs of the Ising model. Nonetheless, numerical experiments 
show that, consistent with the independent case, 
$\rho(\lambda ; L(s) / \sqrt{pN})$ converges weakly 
in distribution to $\rho_{\rm sc} \boxplus \rho_{\rm gauss}$
where $\rho_{\rm gauss}$ is the density of the standard normal distribution and 
the~$\boxplus$ symbol
denotes the \emph{free additive convolution} \cite{WB-AD-TJ:06}. 

\begin{figure}
  \centering
  \begin{figframe}
  \includegraphics[height=0.21\textwidth]{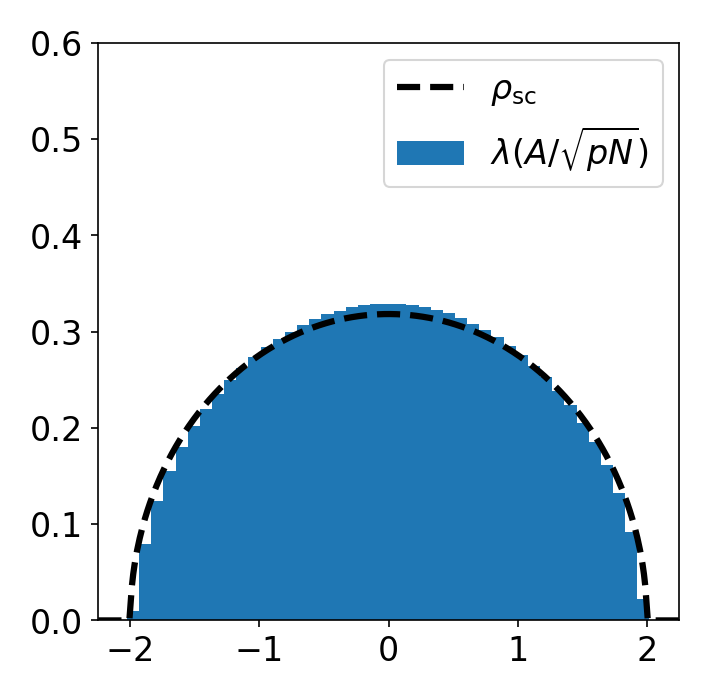}
  \includegraphics[height=0.21\textwidth]{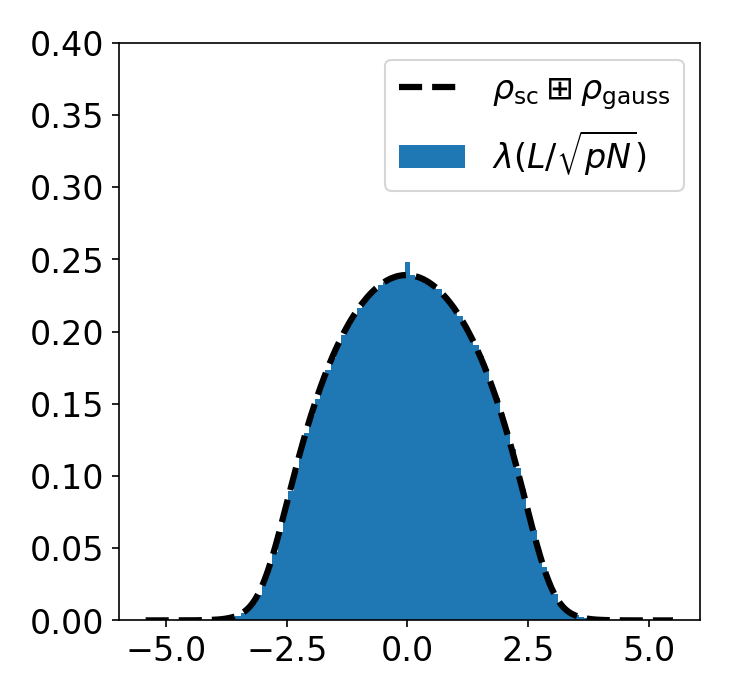}
  \end{figframe}
  \caption{(Left) Histogram of eigenvalues of $\frac{1}{\sqrt{pN}}A(s)$ 
  where $N=1000$ and $p=0.05$ compared against the density of the semi-circle law. 
  (Right) Numerically computed histogram of eigenvalues of 
  $\frac{1}{\sqrt{pN}}L(s)$ where $N=1000$ and $p=0.05$ 
  compared against the density of the free additive 
  convolution of the semi-circle law with the Gaussian distribution (see \cite{WB-AD-TJ:06} for 
  details).}
  \label{fig:esd}
\end{figure}

\begin{figure}
  \centering
  \begin{figframe}
  \includegraphics[height=0.21\textwidth]{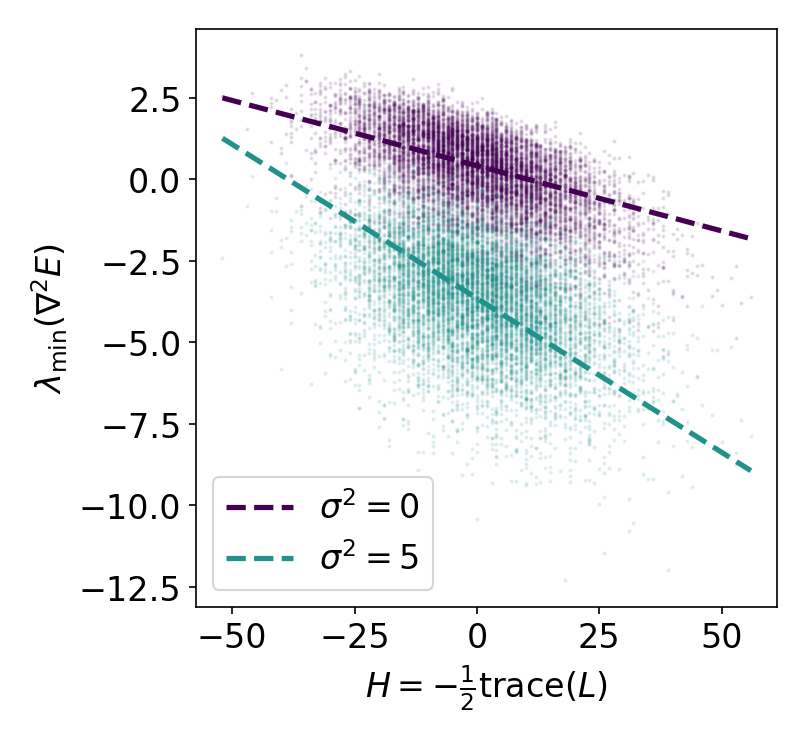}
  \includegraphics[height=0.21\textwidth]{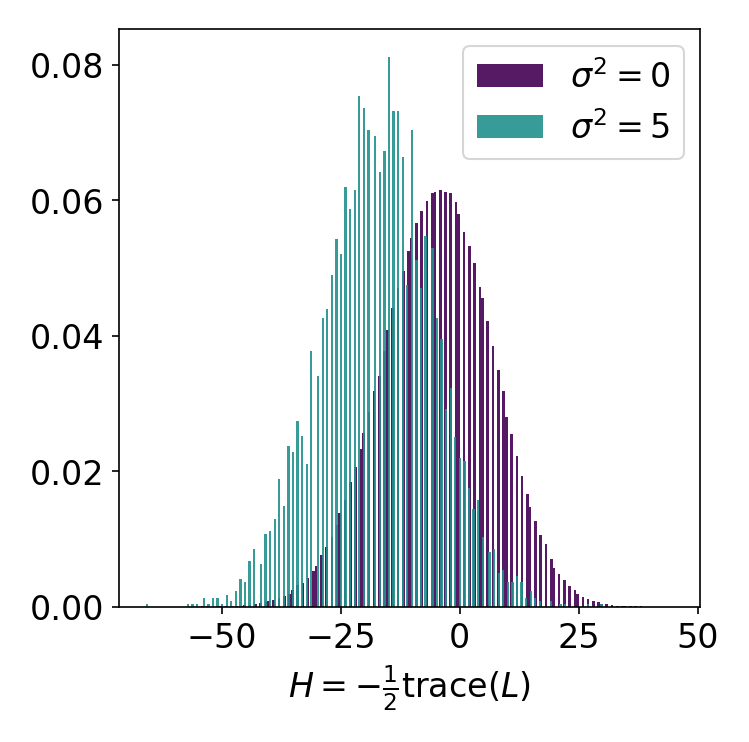}
  \end{figframe}
  \caption{Numerical experiment with $N=50$, $p=0.15$, and $\bar{\mu}=4$. 
  (Left) Plot of minimum 
  Hessian eigenvalue vs. Ising Hamiltonian for 10,000 sampled random graphs.
  The dashed line corresponds to~\eqref{eq:cond-exp}.
  (Right) Histogram of Ising Hamiltonian values corresponding to stable spin configurations.}
  \label{fig:heterogeneity}
\end{figure}

Within this framework, the relationship between the energy and stability of a 
given spin 
configuration can be expressed as the conditional expectation of the minimum 
eigenvalue $\lambda_1(\nabla^2 E_\mu(\phi^*(s)))$ at the energy 
level~$H(s) = h$. This quantity can be approximated as 
\begin{equation}
  \label{eq:cond-exp}
  \begin{aligned}
    &\mathbb{E}\Big[\lambda_1\big(\nabla^2 E_\mu(\phi^*(s))\big) \mid H(s) = h \Big] \\
    &\qquad \approx -\left(\frac{2}{N} + \frac{\sigma^2}{N(N - 1)p}\right)h + 2\bar{\mu} \\
    &\qquad \phantom{=} -\sqrt{\big(2(N - 1)p + 4\sigma^2\big)\log N},
  \end{aligned}
\end{equation}
see~\cite{ANM-AA-FP-AM-etal:26-inprep} for details on the derivation. 
The expression~\eqref{eq:cond-exp} depends inversely on $h$, meaning that as the 
energy of a spin configuration increases, 
on average the minimum eigenvalue decreases, 
and therefore it is less likely to be stable. Moreover, we 
see that increasing heterogeneity of the regularization parameters
by raising $\sigma^2$ both shifts the curve downward and 
steepens its slope cf. Figure~\ref{fig:heterogeneity}(Left). Together, 
these effects selectively stabilize low energy configurations while 
destabilizing high energy configurations, cf. Figure~\ref{fig:heterogeneity}(Right)
where the histogram of Ising Hamiltonian values corresponding to stable 
spin configurations shifts to the left as $\sigma^2$ is increased.

\section{Associative Memory with Coupled Oscillator Networks}
We now discuss the second application of oscillator networks to computing: 
hardware implementations of associative memory. Associative memory refers 
to the ability of a system to store a number of memories, and respond to 
an input stimulus with the memory that most closely resembles it. Mathematical 
models of associative memory are of interest in cognitive neuroscience, 
since they provide insight into mechanisms through which memories are formed and 
recalled in the human brain. Computing applications of associative memory 
models were recognized early leading to interest in hardware implementations 
(see e.g. the \emph{Associatron}~\cite{KN:72}, a 
neural network model implemented on hardware in 1972). These include:

\begin{itemize}
  \item \emph{Pattern Recognition:} Associative memory systems may be used for 
  certain supervised and unsupervised learning tasks involving pattern recognition. 
  For example, if the memories correspond 
  to images, then an input image may be classified using the label 
  of the memorized image an associative memory system produces in response.

  \item \emph{Generative Modeling:} Recent work has identified parallels between 
  associative memory models and diffusion-based generative models~\cite{BH-HS-DK-JH-ZK-DHC:23, LA:24}, 
  and demonstrated that certain associative memory models have the capacity 
  to generalize beyond the memories that they store, enabling the retrieval process 
  to generate new samples. 

  \item \emph{Transformer Architecture:} The attention mechanism, which 
  forms the core of the transformer architecture upon which modern large language models 
  are based, is equivalent 
  to certain associative memory models with the appropriate choice of 
  activation function~\cite{HR-BS-JL-PS-MW-TA-LG-MH-MP-GKS-VG-DK-MK-GK-JB-SH:21-iclr, BH-YL-BP-RP-HS-DHC-MZ-DK:23-neurips}. 
  We extend this connection to the setting of oscillator networks in Section: \sidebarref{sec:attention}. 
\end{itemize}

The most successful model of associative memory is the~\emph{Hopfield model}~\cite{JJH:82, JJH:84}, 
which relates the firing activity of a neuron with a spin variable, and 
encodes memories through the ground states of an Ising Hamiltonian. 
Memories are retrieved using a dynamical system that minimizes this Hamiltonian. 
In the original Hopfield model~\cite{JJH:82}, the dynamics correspond to the discrete-time
McCulloch-Pitts model of a neuron. Various modifications 
have been proposed in both discrete-time and continuous-time that 
generalize the energy function~\cite{MAC-SG:83, JJH:84, DJA-MVT:91, DK-JJH:21-iclr, SB-GB-FB-SZ:25}. 
To facilitate hardware realizations of associative memory, 
in this section we consider models of associative memory where the retrieval dynamics 
take the form of a coupled oscillator network.

\subsection{The Hopfield Model}
\begin{sidebarfixed}{Hebbian Learning in Oscillator Networks}
  \label{sidebar:hebbian}
  \sdbarinitial{H}ebbian learning is a class of unsupervised learning rules for neural 
  networks inspired by a theory of learning in neurobiological systems 
  proposed by D. O. Hebb, often stated as the aphorism ``neurons 
  that fire together, wire together''. Here, learning can be viewed as 
  a dynamical system governing the evolution of the weights $J_{ij}$ of a network 
  which depends on the oscillatory phases
  \begin{sequation}
    \label{seq:generalized-hebbian-learning}
    \dot{J_{\phantom{ij}}}\!\! = f(J; \phi).
  \end{sequation}

  Mathematical formulations of Hebbian learning~\cite{WG-WMK:02, FB:25-lnd} typically 
  require~\eqref{seq:generalized-hebbian-learning} to satisfy, 
  among other conditions, the following criteria
  \begin{enumerate}
    \item \textbf{Locality}: The evolution of the weight of the 
    edge $(i, j)$ depends only on the states $\phi_i$ and $\phi_j$,
      $\dot{J_{ij}} = f_{ij}(J_{ij}, \phi_i, \phi_j)$
    \item \textbf{Cooperativity}: The weight of an edge $J_{ij}$ increases 
    when the phases of the corresponding nodes, $\phi_i$ and $\phi_j$, are synchronized. 
    \item \textbf{Synaptic Depression}: The weight $J_{ij}$ decays 
    in the absence of co-activity of $i$ and $j$.  
  \end{enumerate}

  In the context of oscillator networks, a candidate satisfying these conditions is 
  \[ f_{ij}(J_{ij}, \phi_i, \phi_j) = -J_{ij} + \cos(\phi_i - \phi_j). \]
  The weight assignments for associative memory systems~\eqref{eq:Hebbian} and~\eqref{eq:phase-hebbian} 
  can be viewed as a steady-state of this rule. 

  Typically, learning takes place on a slower timescale 
  than that of the phase dynamics of the oscillator network. 
  Thus, the combination of the two is a multiple timescale system 
  \begin{sequation}
    \label{seq:coupled-hebbian-system}
    \begin{aligned}
      \tau\dot{\phi}_i &= \omega_i + \sum_{j=1}^{N}J_{ij}\sin(\phi_j - \phi_i) \\
      \dot{J_{ij}} &= f_{ij}(J_{ij}; \phi_i, \phi_j).
    \end{aligned}
  \end{sequation}
  where $\tau \ll 1$. Such systems exhibit a rich variety of synchronous 
  behavior and have been extensively 
  studied in~\cite{PS-SCY-LST:02, QR-JZ:07, TA-TA:09, RKN-LQE:09, RB-ES-SY:19}. 

  Beyond biological modeling, 
  Hebbian learning rules are also of importance in machine learning. 
  For instance, neural networks are sometimes trained using 
  \emph{contrastive Hebbian learning}~\cite{DHA-GEH-TJS:85, JRM:91}
  rules of the form 
  \begin{sequation}
    \label{seq:contrastive-hebbian}
    \dot{J_{ij}} = \cos(\phi_i - \phi_j) - \cos(\tilde{\phi}_i - \tilde{\phi}_j)
  \end{sequation}
  where $\phi$ denotes the phases of the network when evolving freely, 
  and $\tilde{\phi}$ denotes the phases of the network when the outputs are clamped 
  at a target value. Related learning rules arise in training algorithms for 
  energy models such as equilibrium propagation~\cite{BS-YB:17}, where a contrastive 
  update, similar to~\eqref{seq:contrastive-hebbian}, estimates the gradient of a loss 
  function. Unlike backpropagation, rules of this form are suitable 
  for implementation in hardware since the weight updates can be computed 
  using the dynamics of the oscillator network itself, as
  demonstrated in~\cite{JL-DM-JG:24, QW-CCW-FM:24, TR-JG:25, AG:25}.
  \putbib[refs]
\end{sidebarfixed}

\begin{figure*}
  \makebox[\textwidth][c]{
  \begin{subfigure}[b]{0.105\textwidth}
    \centering
    \includegraphics[width=\linewidth, trim=46 39 37 43, clip]{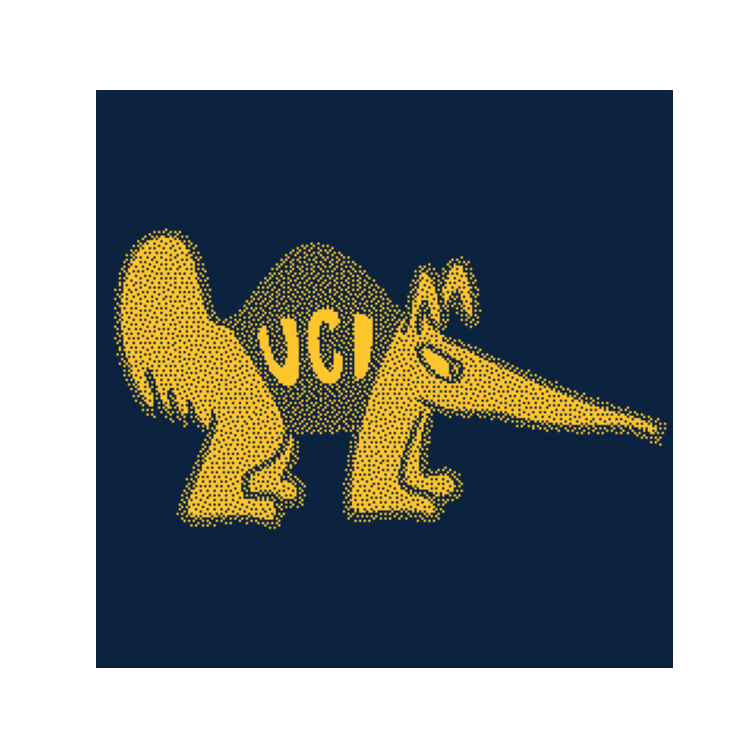}\\[2pt]
    \includegraphics[width=\linewidth, trim=46 39 37 43, clip]{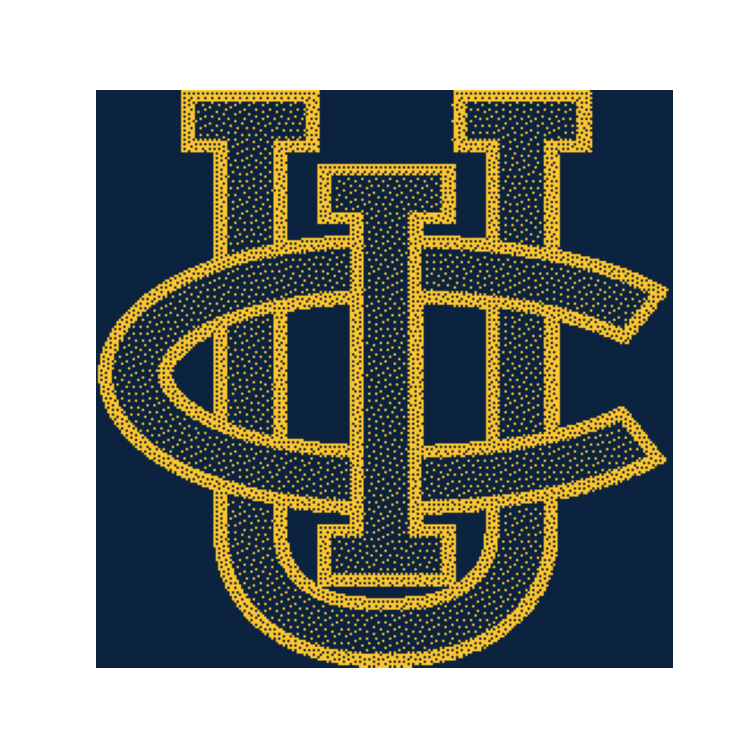}
  \end{subfigure}
  \hfill
  \includegraphics[height=0.215\textwidth]{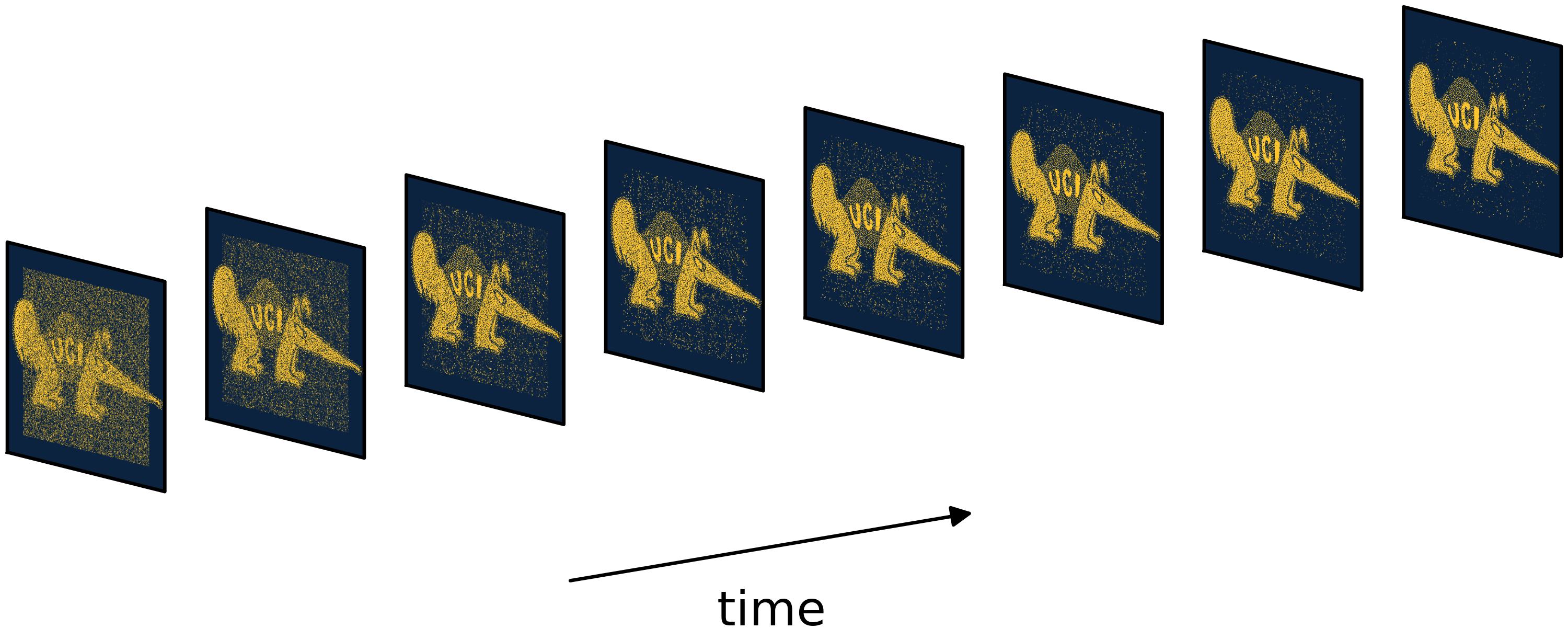}
  \hfill
  \includegraphics[height=0.215\textwidth, trim=46 39 37 43, clip]{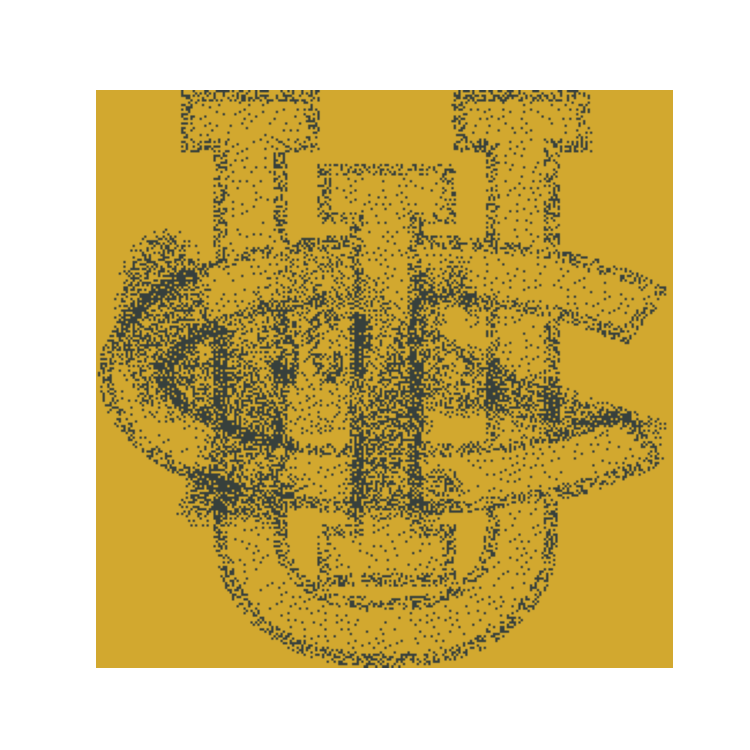}
  }
  \caption{Encoding images in the attractors of an associative memory system~\eqref{eq:kuramoto-memory-dynamics}. Each 
  node corresponds to a pixel in a $N_{\rm grid} \times N_{\rm grid}$ bitmap, 
  where $N_{\rm grid} = 250$. The color of each pixel is equal to the phase of the corresponding node. 
  (Left) Two memories stored in the network using the Hebbian learning rule~\eqref{eq:phase-hebbian}. 
  (Center) Solution of~\eqref{eq:kuramoto-memory-dynamics} starting from a corrupted version of the second memory. The 
  system successfully recovers the correct memory. (Right) Spurious attractor of the network.}
  \label{fig:memory-images}
\end{figure*}

We begin by formalizing the associative memory problem and reviewing 
the Hopfield model (we refer 
the reader to~\cite{JH-AK-RGP:91} for a detailed introduction to the Hopfield 
model and its extensions). 
Suppose 
that we have a set of $M$ memories or patterns
represented by spin configurations, denoted by~$\xi^{(1)}, \xi^{(2)}, \dots, \xi^{(M)} \in \{-1, 1\}^N$. 
We wish to design a dynamical system that, when given a new pattern 
$\zeta \in \{-1, 1 \}^N$ specifying its initial condition,
the state converges to the stored memory $\xi^{(i)}$ that is closest to $\zeta$.

In the Hopfield model, this problem is solved by choosing the weights $J_{ij}$
of an Ising Hamiltonian~\eqref{eq:ising-hamiltonian} (with field $h_i = 0$) so that its ground 
states correspond to the memories we would like to store. 
The weights are chosen according to the \emph{Hebbian learning rule}, 
\begin{equation}
  \label{eq:Hebbian}
  J_{ij} = \frac{1}{N}\sum_{\mu=1}^{M}\xi^{(\mu)}_i\xi^{(\mu)}_j,
\end{equation}
which acquires its name from a hypothesis 
by Hebb~\cite{DOH:49} that changes in the connection strength between two 
neurons $i$ and $j$ should be proportional to the correlation in their firing
activity, which in this case
is represented by the product~$\xi^{(\mu)}_i\xi^{(\mu)}_j$, 
see also Sidebar: \sidebarref{sidebar:hebbian}. 

To minimize the Ising Hamiltonian, Hopfield introduces the 
following dynamics, based on the McCulloch-Pitts model
of a neuron~\cite{WSM-WP:43},
\begin{equation}
  \label{eq:Hopfield-dynamics}
  s_{i}^{+} = \operatorname{sgn}\left(\sum_{j=1}^{N}J_{ij}s_j\right).
\end{equation}
It is not necessarily the case that the memories 
correspond to stable equilibria of~\eqref{eq:Hopfield-dynamics}. If 
the number of memories is too large, relative to the size of the network, 
then the learning rule~\eqref{eq:Hopfield-dynamics} may instead introduce spurious 
minima unrelated to the desired memories. A central concern in the study of 
such neural networks is their \emph{storage capacity}, i.e. the maximum number 
of memories that can be stored as attractors of~\eqref{eq:Hopfield-dynamics}. 

Estimates of the storage capacity differ depending on whether one allows for small 
errors in the retrieved memories. We consider both cases here.  
Assume that the memories~$\xi^{(1)}, \xi^{(2)}, \dots, \xi^{(M)}$ 
are sampled independently and uniformly at random from the set 
of all possible spin configurations~$\{-1, 1\}^N$. Then,

\begin{itemize}
  \item If small errors are permitted, then as $N \to \infty$, the maximum number 
  of storable memories $M_{\rm max}$ satisfies $\alpha_{\rm max} = \frac{M_{\rm max}}{N} \to 0.138$. 
  \item If errors are not permitted, the maximum number of storable 
  memories is $M_{\rm max} \approx \frac{N}{4\log N}$ when $N$ is large. 
\end{itemize}

\subsection{Oscillatory Associative Memory}
We now discuss an oscillatory associative memory system based 
on the standard Kuramoto model~\cite{AA-CJPV:94, TA:98}. This model generalizes 
beyond binary-valued memory vectors to those represented by continuous 
phase values, which we denote
$\varphi^{(1)}, \varphi^{(2)}, \dots, \varphi^{(M)} \in \mathbb{T}^N$. 
The goal is to design a coupled oscillator network that stores 
these phase configurations as attractors such that the network converges to the 
memory nearest its initial phase configuration. 

The energy function is 
\begin{equation}
  \label{eq:kuramoto-memory-energy}
  E(\phi) = -\frac{1}{2}\sum_{i=1}^{N}\sum_{j=1}^{N}J_{ij}\cos(\phi_j - \phi_i).
\end{equation}
As with the Hopfield model, we wish to encode the memories as the minimizers of~\eqref{eq:kuramoto-memory-energy}.
We accomplish this by applying a learning rule analogous to~\eqref{eq:Hebbian}.
\begin{equation}
  \label{eq:phase-hebbian}
  J_{ij} = \frac{1}{N}\sum_{\mu=1}^{M}\cos(\varphi^{(\mu)}_i - \varphi^{(\mu)}_j).
\end{equation}
Similar to the standard Hebbian learning rule, the connection strength between 
nodes $i$ and $j$ is equal to the average correlation between their phases, 
represented by the quantity $\cos(\varphi^{(\mu)}_i - \varphi^{(\mu)}_j)$. 

To compute the ground states of the energy function, we employ the gradient dynamics, 
which in this case results in the Kuramoto model, 
\begin{equation}
  \label{eq:kuramoto-memory-dynamics}
  \dot{\phi}_i = \sum_{j=1}^{N}J_{ij}\sin(\phi_j - \phi_i)
\end{equation}
Certain works~\cite{AA-CJPV:94, CJPV-AA-LLB:96, MY-MS-MY:99, MY-MS:00} have also considered 
the effect of white noise, delays, and heterogeneity in the natural frequencies 
of oscillators on storage capacity and retrieval accuracy, but for simplicity 
we focus on the standard Kuramoto formulation. 

We demonstrate the on an example where we would like to encode images 
in the attractors of~\eqref{eq:kuramoto-memory-dynamics}. Consider 
a network where each node corresponds to a pixel in a $N_{\rm grid} \times N_{\rm grid}$ 
bitmap, and the value of each pixel is given by the phase of its node. 
In Figure~\ref{fig:memory-images} we demonstrate an example where two 
images are learned in a network with $N_{\rm grid} = 250$. The solution 
of~\eqref{eq:kuramoto-memory-dynamics} starting from a memory corrupted 
with noise is shown in Figure~\ref{fig:memory-images}(Center), and we 
see the system successfully removes the noise and retrieves the 
correct memory. However, the system also has spurious attractors, shown 
in Figure~\ref{fig:memory-images}(Right). 

The stability of the memory $\varphi^{(\mu)}$ can be analyzed by introducing 
a quantity $m^{(\mu)} \in \complex$ called the \emph{overlap} where
\[ m_\mu = \frac{1}{N}\sum_{i=1}^{N}e^{\mathrm{i}(\phi_i - \varphi^{(\mu)}_i)} \]
The magnitude $R_\mu = |m_\mu|$ of the overlap measures the closeness of the 
state $\phi$ to the memory $\varphi^{(\mu)}$ up to a constant phase shift 
$\psi_\mu = \angle m_\mu$.
If $R_\mu = 1$, then $\phi_i = \varphi^{(\mu)}_i + \psi_\mu$ for 
all $i=1, 2, \dots, N$, whereas if $R_\mu\approx 0$, then the phases are incoherent and the 
system has failed to retrieve the memory $\varphi^{(\mu)}_i$. 
Thus the storage capacity and the retrieval accuracy can be estimated by studying the asymptotic 
behavior of the order parameters when the dynamics~\eqref{eq:kuramoto-memory-dynamics} 
are initialized near the memory $\varphi^{(\mu)}$. 

\begin{figure}
  \centering
  \includegraphics[width=0.4\textwidth]{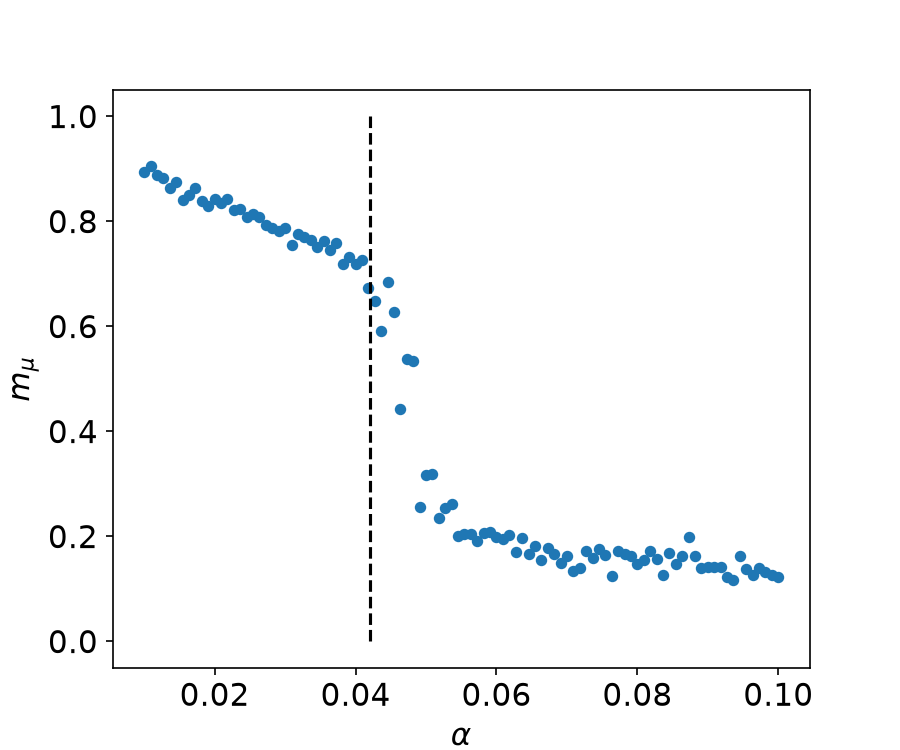}
  \caption{Overlap vs. memory density for associative memory model~\eqref{eq:kuramoto-memory-dynamics}
  with $N=2000$ nodes. The dashed line corresponds to the theoretical memory 
  capacity $\alpha_{\rm max} = 0.042$.}
  \label{fig:overlap}
\end{figure}

Analysis of the storage capacity of the Kuramoto model was performed in~\cite{TA:98}.
In the limit as~$N \to \infty$ the storage capacity 
is~$\alpha_{\rm max} = \frac{M_{\rm max}}{N} \to 0.042$, which is approximately $1/3$ that of 
the classical Hopfield model. With the model~\eqref{eq:kuramoto-memory-dynamics}, 
the retrieval process will inevitably introduce errors, and in fact the exact memory 
patterns are unstable whenever~$M > 2$. In Figure~\ref{fig:overlap} we plot 
the average overlap for various values of the memory density $\alpha = \frac{M}{N}$
on a network with $N=2000$ nodes. We see that the network is capable of 
accurately retrieving memories whenever $\alpha < \alpha_{\rm max}$, 
and the accuracy degrades once $\alpha \geq \alpha_{\rm max}$. 

\subsection{Oscillatory Associative Memory with Error-free Retrieval}
We now discuss a modification to the Kuramoto associative memory model 
to allow for error-free retrieval~\cite{TN-YL-FCH:04, TN-FCH-YL:04}. In 
this formulation we return to the setting where the memories 
are binary-valued, and we again denote them by 
$\xi^{(1)}, \xi^{(2)}, \dots, \xi^{(M)} \in \{-1, 1\}^N$. 

To derive the error-free dynamics, we consider the following energy 
function which modifies~\eqref{eq:kuramoto-memory-energy} with an 
additional all-to-all second order coupling term,
\begin{equation}
  \label{eq:kuramoto-nishikawa-energy}
  \begin{aligned}
    E_\epsilon(\phi) = &-\frac{1}{2} \sum_{i=1}^{N}\sum_{j=1}^{N}J_{ij}\cos(\phi_j - \phi_i) \\ 
      &-\frac{\epsilon}{4N}\sum_{i=1}^{N}\sum_{j=1}^{N}\cos(2(\phi_j - \phi_i))
  \end{aligned}
\end{equation}
The network 
weights are chosen according to the standard Hebbian learning 
rule~\eqref{eq:Hebbian}. Taking the gradient of~\eqref{eq:kuramoto-nishikawa-energy}
results in the retrieval dynamics 
\begin{equation}
  \label{eq:kuramoto-nishikawa}
  \dot{\phi}_i = \sum_{j=1}^{N}J_{ij}\sin(\phi_j - \phi_i) + \frac{\epsilon}{N}\sum_{j=1}^{N}\sin(2(\theta_j - \theta_i))
\end{equation}

\begin{figure}
  \centering
  \includegraphics[width=0.4\textwidth]{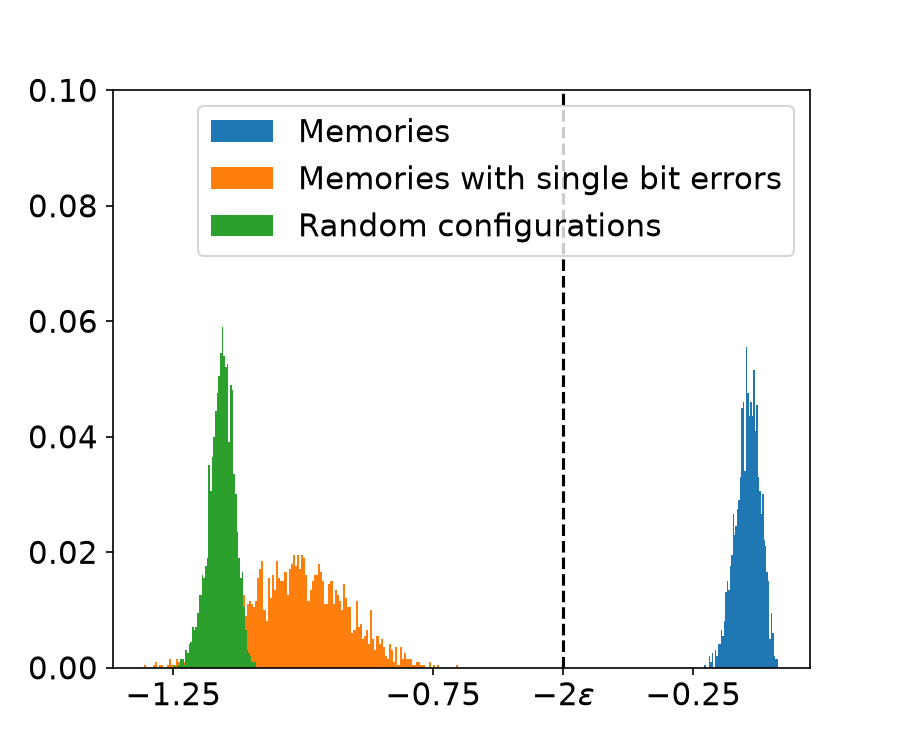}
  \caption{Histogram of $\lambda_{1}(L(s))$ for randomly sampled networks with $N=1000$ nodes and $M=10$ memories, 
  where $s$ corresponds to a memory, a memory with a random single bit error, or a spin configuration sampled uniformly 
  at random from $\{-1, 1\}^N$. Only memories satisfy the stability condition $\lambda_{1}(L(s)) > -2\epsilon$.}
  \label{fig:nishikawa-hist}
\end{figure}

For every spin configuration $s \in \{-1, 1\}^N$, the system~\eqref{eq:kuramoto-nishikawa}
has a unique equilibrium~$\phi^*(s)$ with the property that
\[ 
  \phi_i^*(s) - \phi_j^*(s) = \begin{cases}
    0 \qquad s_i = s_j \\ 
    \pi \qquad s_i \neq s_j
  \end{cases}
\]
As we show next, the $\sin(2(\theta_j - \theta_i))$ coupling term plays a similar 
role to second harmonic injection locking Oscillator Ising machines, and preferentially 
stabilizes equilibria corresponding to memories,~$\phi^*(\xi^{(\mu)})$, while 
leaving arbitrary patterns unstable. 
Since~\eqref{eq:kuramoto-nishikawa} is a gradient system, the stability of 
an equilibrium $\phi^*(s)$ can be established by inspecting the Hessian 
of the energy function~\eqref{eq:kuramoto-nishikawa-energy}. 
\[ \nabla^2 E_\epsilon(\phi^*(s)) = L(s) + 2\epsilon I - \frac{2\epsilon}{N}\mathbb{1}_{N \times N} \]
where $L(s)$ is the signed graph Laplacian~\eqref{eq:signed-laplacian}, 
and $\mathbb{1}_{N \times N}$ is the $N \times N$ matrix of ones. By 
\cite[Theorem 2]{TN-FCH-YL:04} the equilibrium 
$\phi^*(s)$ is stable if $\lambda_{1}(L(s)) > -2\epsilon$, and 
unstable if $\lambda_{1}(L(s)) < -2\epsilon$, where $\lambda_1(L(s))$ denotes 
the minimum eigenvalue of $L(s)$. 

To characterize the stability of memories and the accuracy of the memories 
retrieved by the system \eqref{eq:kuramoto-nishikawa}, we sample 
networks of the form \eqref{eq:Hebbian} where $M=10$ and the memories 
are chosen uniformly at random. In Figure~\ref{fig:nishikawa-hist}, we plot 
the histograms of the minimum Laplacian eigenvalues $\lambda_{1}(L(s))$,  
where $s$ is either a memory, a memory with a randomly introduced 
bit error, or a spin configuration sampled uniformly at random from $\{-1, 1\}^N$. 
Eigenvalues corresponding to memories are significantly larger 
than those corresponding to either corrupted memories or arbitrary configurations. 
If $\epsilon = 0.25$, the set of stable equilibria will consist almost exclusively 
of memories, in which case the system~\eqref{eq:kuramoto-nishikawa} is capable of 
nearly error-free retrieval. 

The error-free memory capacity \cite[Theorem 5]{TN-FCH-YL:04} of this model 
is $M_{\rm max} = \frac{2\epsilon^2 N}{\log N}$. Increasing $\epsilon$
raises the memory capacity of the network, as well as the stability margin of equilibria that correspond to memories. 
However, if $\epsilon$ is too large, this may also stabilize corrupted memories and other 
arbitrary spin configurations. For constant $\epsilon$, the memory capacity scales 
asymptotically at the same rate as that of the classical Hopfield model. 

\section{Attention with Coupled Oscillator Networks}
\label{sec:attention}
We discuss here the final application of oscillator networks to computing: implementing 
the attention mechanism, a type of neural network layer
that is a fundamental component of the transformer architecture on which modern large 
language models (LLMs) are based. The attention mechanism operates on sequences by 
computing a measure of the importance of a position within an input sequence to 
a position in the output sequence \cite{AV-NS-NP-JU-LJ-ANG-LK-IP:17-neurips, DB-KC-YB:15-iclr}. 
This correlation measure is expressed in terms of the similarity between encodings of the input 
sequence called keys and queries. Since the values corresponding to different positions can be 
computed in parallel, the transformer replaced recurrent neural network architectures, 
which are inherently sequential, as the dominant approach to language modeling. 

Transformer-based LLMs have proven to be extremely effective, enabling 
an explosion in artificial intelligence capabilities since their introduction. 
However, their biggest limitation is the massive amount of computational resources 
they require due to the enormous scale of AI models. Here we show that the computation 
performed by the attention mechanism is equivalent to the dynamics of certain higher-dimensional 
coupled oscillator networks. This paves the way for hardware implementations of the
attention mechanism through synchronizing oscillators, a first step in the development of a new 
class of oscillator-based AI systems. 

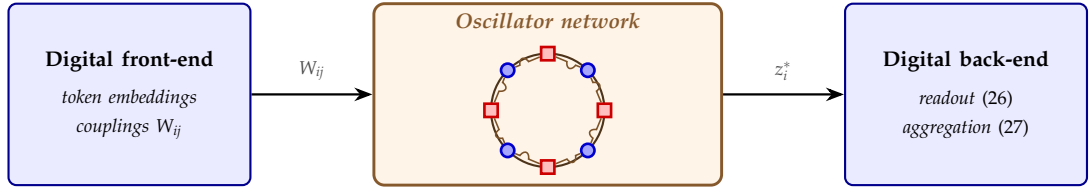
\begin{figure*}
\centering
\begin{tikzpicture}[
  font=\small, >=Stealth,
  digital/.style={rectangle, rounded corners=3pt,
    draw=blue!55!black, line width=0.8pt, fill=blue!8,
    minimum width=3.2cm, minimum height=2.4cm,
    align=center, inner sep=5pt},
  analog/.style={rectangle, rounded corners=4pt,
    draw=brown!70!black, line width=1.1pt, fill=orange!9,
    minimum width=4.6cm, minimum height=2.4cm,
    align=center, inner sep=5pt},
  keyosc/.style={circle, draw=blue!80!black, line width=1.0pt,
    fill=blue!35, minimum size=5pt, inner sep=0pt},
  queryosc/.style={rectangle, draw=red!80!black, line width=1.0pt,
    fill=red!25, minimum size=5pt, inner sep=0pt},
  spring/.style={decorate,
    decoration={coil, aspect=0.5, segment length=3pt,
      amplitude=2pt, pre length=2pt, post length=2pt},
    draw=brown!65!black, line width=0.6pt},
  arrowlabel/.style={font=\scriptsize\itshape, text=black!70},
]
\node[digital] (frontend) at (0,0)
  {\textbf{Digital front-end}\\[3pt]
   \scriptsize\itshape token embeddings\\
   \scriptsize\itshape couplings $W_{ij}$};

\node[analog, right=1.6cm of frontend] (analog)
  {\phantom{X}\\[28pt]\phantom{X}};
\node[anchor=north, font=\small\bfseries\itshape,
      text=brown!70!black, inner sep=0pt]
  at ([yshift=-4pt]analog.north) {Oscillator network};

\coordinate (ctr) at ([yshift=-6pt]analog.center);
\draw[brown!40!black, line width=0.8pt] (ctr) circle (.75cm);
\foreach \j/\ang in {0/90,1/180,2/270,3/0}
  \node[queryosc] (q\j) at ($(ctr)+(\ang:.75cm)$) {};
\foreach \i/\ang in {0/45,1/135,2/225,3/315}
  \node[keyosc]   (k\i) at ($(ctr)+(\ang:.75cm)$) {};
\draw[spring](k0)--(q0); \draw[spring](k0)--(q3);
\draw[spring](k1)--(q0); \draw[spring](k1)--(q1);
\draw[spring](k2)--(q1); \draw[spring](k2)--(q2);
\draw[spring](k3)--(q2); \draw[spring](k3)--(q3);

\node[digital, right=1.6cm of analog] (backend)
  {\textbf{Digital back-end}\\[3pt]
   \scriptsize\itshape readout~\eqref{eq:lohe-attention-matrix}\\
   \scriptsize\itshape aggregation~\eqref{eq:lohe-output}};

\draw[->, line width=0.85pt] (frontend.east) -- (analog.west)
  node[midway, above=2pt, arrowlabel] {$W_{ij}$};
\draw[->, line width=0.85pt] (analog.east) -- (backend.west)
  node[midway, above=2pt, arrowlabel] {$z^*_i$};

\end{tikzpicture}
\caption{Architecture of oscillator-based implementation of the attention mechanism. A digital 
front-end provides the input sequence and weights $W_{ij}$. Once the oscillator network settles 
to equilibrium, the digital back-end reads the attention matrix~\eqref{eq:lohe-attention-matrix}, 
and computes the output sequence~\eqref{eq:lohe-output}.}
\label{fig:architecture}
\end{figure*}

\subsection{Self-Attention}
We begin by reviewing self-attention in modern transformers. 
Let $x_1, x_2, \dots, x_T \in \real^{d_{\rm in}}$ be a sequence of vectors, called 
\emph{tokens}, each of which encodes a symbol. Given matrices 
$W_K \in \real^{d \times d_{\rm in}}$, $W_Q \in \real^{d \times d_{\rm in}}$, and 
$W_V \in \real^{d \times d_{\rm out}}$, which are learned from data,
let $k_i = W_K x_i$, $q_i = W_Q x_i$, and $v_i = W_V x_i$ be the \emph{keys}, \emph{queries},
and \emph{values} respectively.

In the transformer architecture, the self-attention layer computes an output 
sequence~$y_1, y_2, \dots, y_T \in \real^{d_{\rm out}}$ 
in terms of the values, and matrices of queries 
$Q = [q_1 \cdots q_T]$ and keys $K = [k_1 \cdots k_T]$, 
where
\[ y_i = \sum_{j=1}^{T}\alpha_{ij}(Q, K)v_j \qquad i=1, \dots, T  \]
and the matrix $A = (\alpha_{ij}(Q, K))_{i,j=1}^{T}$ is 
called the \emph{attention matrix}. Over the years, various attention 
matrices have been proposed, including
\begin{itemize}
  \item \emph{Softmax attention} \cite{AV-NS-NP-JU-LJ-ANG-LK-IP:17-neurips, DB-KC-YB:15-iclr}:
  \[ \alpha_{ij}(Q, K) = \frac{\exp(\frac{q_i^\top k_j}{\sqrt{d}})}{\sum_{\ell=1}^{T}\exp(\frac{q_i^\top k_\ell}{\sqrt{d}})}. \]
  \item \emph{Sigmoid attention} \cite{JR-FD-EGD-FW-DB-PA-TL-JD-ZG-AS-RW:25-iclr}:
  \[ \alpha_{ij}(Q, K) = \frac{1}{1 + \exp(-\frac{q_i^\top k_j}{\sqrt{d}})} \]
  \item \emph{Linear attention} \cite{AK-AV-NP-FF:20-icml}:
  \[ \alpha_{ij}(Q, K) = \frac{\varphi(q_i)^\top \varphi(k_j)}{\sum_{\ell=1}^{T}\varphi(q_i)^\top \varphi(k_\ell)}, \]
  where $\varphi:\real^{d} \to \real^{d}$ is a nonlinear feature map.
\end{itemize}

In each case, the weight $\alpha_{ij}(Q, K)$ is interpreted as the importance 
of the~$j$th token of the input sequence to the~$i$th token of the output 
sequence, where the importance corresponds to the relative magnitude of the 
inner product~$q_i^\top k_j$. In other words, the more the~$j$th key 
is aligned with the~$i$th query, relative to the keys corresponding to other positions 
in the sequence, the more the~$j$th input token influences the~$i$th output token. 

\subsection{Self-Attention via Coupled Oscillator Networks}
We now discuss an implementation of the attention mechanism 
as a higher-dimensional coupled oscillator network proposed 
in~\cite{FP-TG:26}. The architecture of the system 
is shown in Figure~\ref{fig:architecture}, and is derived 
by drawing an analogy between the alignment of keys and
queries, and the synchronization of phases in an oscillator 
network. 

Consider a bipartite 
network with $T$ \emph{free oscillators} denoted 
$z_1, z_2, \dots, z_T \in S^{d - 1}$, 
and $T$ \emph{anchors} denoted 
$r_1, r_2, \dots, r_T \in S^{d - 1}$. The values 
of the anchors are considered learned parameters. 
Let $W_{ij} = \sigma(\beta q_i^\top k_j)$
denote the weight of the connecting edge 
between the $i$th free oscillator and the $j$th anchor, 
where $\sigma:\real \to \real$ is a sigmoid function, 
and $\beta$ is an inverse temperature parameter. 

In the model, the anchors remain fixed while the 
free oscillators evolve according to
\begin{equation}
  \label{eq:free-oscillators}
  \dot{z}_i = (I - z_iz_i^\top)\sum_{j=1}^{T}W_{ij}r_j
\end{equation} 
This system is related to multi-agent formulations 
of transformers considered 
in~\cite{YL-ZL-DH-ZS-BD-TQ-LW-TL:19, BG-CL-YP-PR:23-neurips, BG-CL-YP-PR:25, CA:26-tac}
except that in~\eqref{eq:free-oscillators} the weights are 
static rather than state-dependent, 
making it more amenable to hardware 
implementation. 

The system~\eqref{eq:free-oscillators} is a special 
case of the high-dimensional Kuramoto model~\eqref{eq:kuramoto-sphere}, 
which is the projected gradient flow of 
the Hamiltonian of the $d$-vector spin glass model~\eqref{eq:d-vector-hamiltonian},
where the free oscillators are non-interacting, i.e. $J_{ij} = 0$, 
and the external field is  
\begin{equation}
  \label{eq:attention-field}
  h_i = \sum_{j=1}^{T}W_{ij}r_j.
\end{equation}
In other words,~\eqref{eq:free-oscillators} is of the form
$\dot{z}_i = \Pi_{T_{z_i}S^{d - 1}}(-\nabla_{z_i}H(z))$,
where $\Pi_{T_{z_i}S^{d - 1}} = I - z_iz_i^\top$ denotes 
the orthogonal projection onto the tangent space $T_{z_i}S^{d - 1}$. The projection 
ensures that the states of the free oscillators remain on the unit
sphere~$S^{d-1}$ for all time. 
The system has exactly two equilibria, $z^*$ and~$-z^*$, where
\[ z_i^* = \frac{h_i}{\norm{h_i}}.  \]
Using the Hamiltonian~\eqref{eq:d-vector-hamiltonian} as a Lyapunov function, 
it can be shown that the only stable equilibrium is $z_i^*$, with basin of 
attraction $S^{d - 1} \setminus \{-z_i^* \}$. Thus if the states
of~\eqref{eq:free-oscillators} are initialized randomly, the system 
will almost surely converge to the equilibrium $z^*$. 

Once the system has converged to a steady state, 
the attention matrix is built as 
\begin{equation}
  \label{eq:lohe-attention-matrix}
  a_{ij} = \frac{1 + (z_i^*)^\top r_j}{1 + \sum_{k=1}^{N}(z_i^*)^\top r_k} 
\end{equation}
and the output sequence is 
\begin{equation}
  \label{eq:lohe-output}
  y_i = \sum_{j=1}^{T}a_{ij}v_j
\end{equation}
In practical implementations, the quantity~\eqref{eq:lohe-attention-matrix}
is not computed but rather obtained via a measurement of the state using 
analog hardware. For example, in the special case where $d = 2$, the states of the free 
oscillators and anchors can be written as 
$z_i^* = [\cos(\phi_i^*), \sin(\phi_i^*)]$, and $r_j = [\cos(\varphi_j), \sin(\varphi_j)]$, 
and the inner product can be expressed entirely in terms of the 
phase difference of oscillations $(z_i^*)^\top r_j = \cos(\phi_i^* - \varphi_j)$, 
which can be obtained via measurements of the steady-state oscillations.  

\begin{figure}
  \centering
  \includegraphics[width=0.39\textwidth]{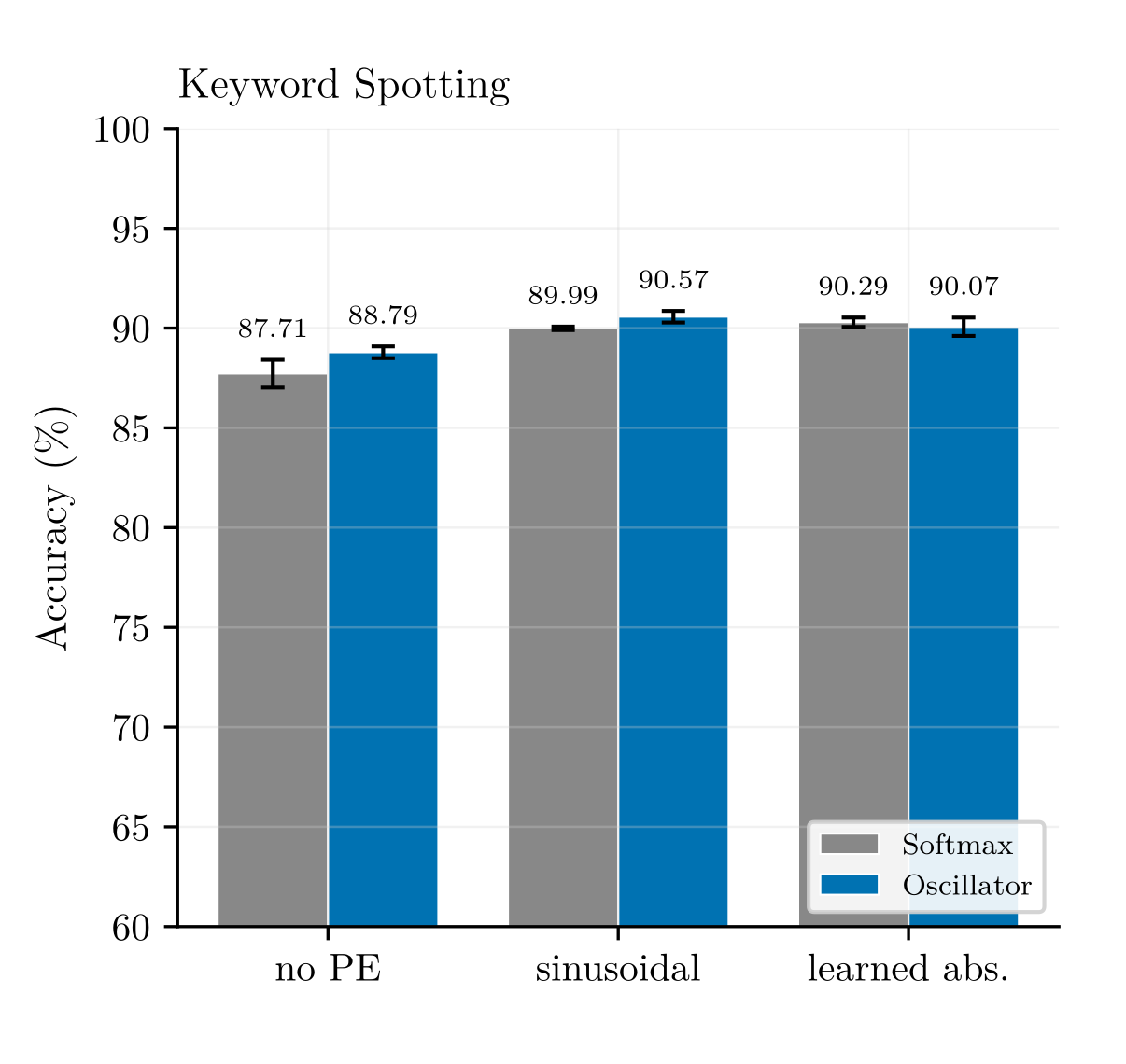}
  \includegraphics[width=0.39\textwidth]{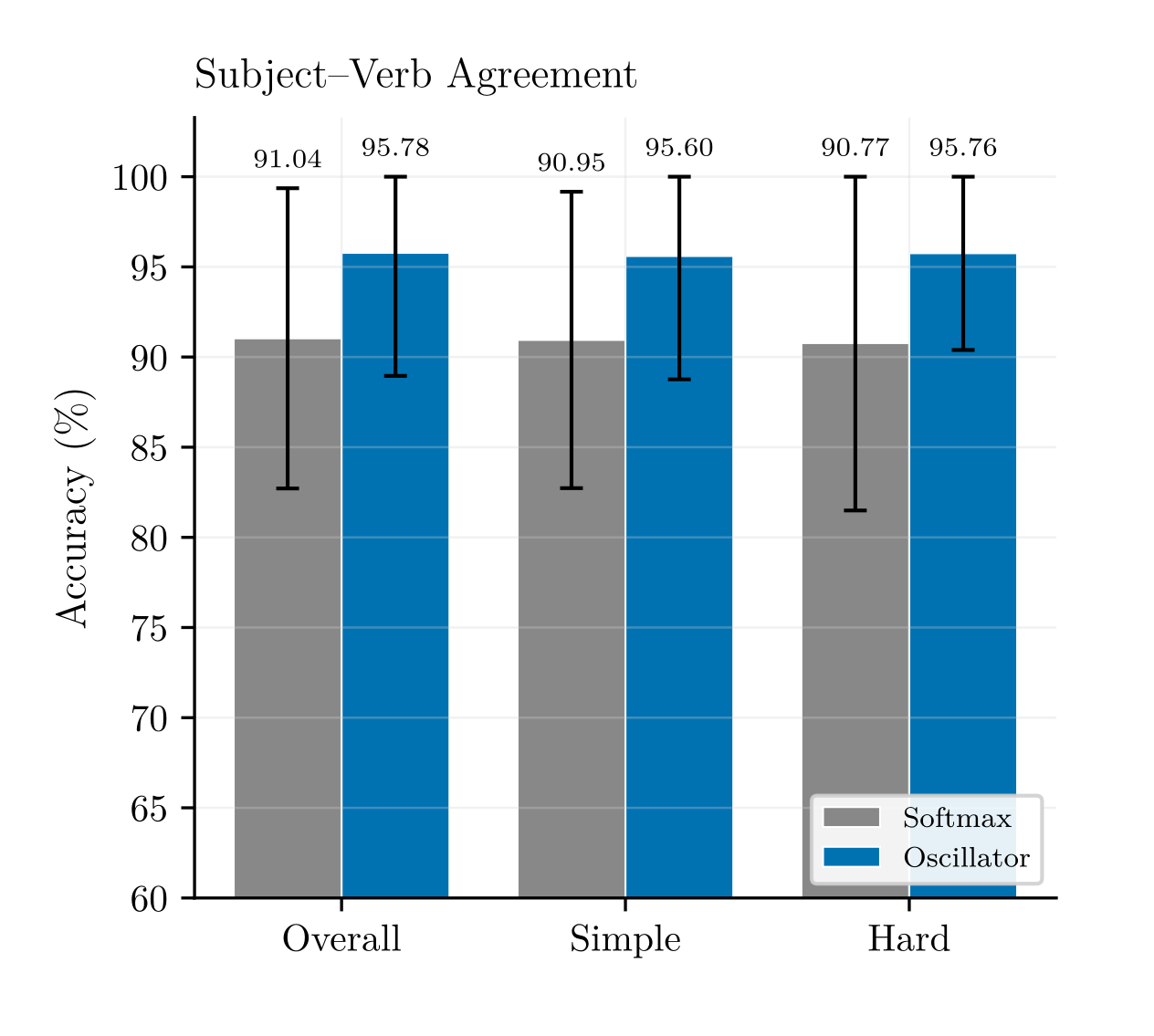}
  \caption{Comparison between softmax attention and oscillator-based attention with $d=32$ 
  on performance on benchmark tasks with different types of 
  positional encoding of input sequences. (Top) Accuracy on keyword spotting task (Bottom) Accuracy 
  of subject-verb agreement of predicted text.}
  \label{fig:benchmark}
\end{figure}

In benchmark experiments~\cite{FP-TG:26}, oscillator-based attention achieves 
similar performance to the standard softmax attention 
in~\cite{AV-NS-NP-JU-LJ-ANG-LK-IP:17-neurips}. Figure~\ref{fig:benchmark}
compares the accuracy of the oscillator-based 
attention mechanism to the softmax attention on a keyword spotting task and 
subject-verb agreement task respectively. Though the classical attention has a 
slight advantage, experimental results suggest that hardware implementations via 
oscillator networks are a viable alternative. 

We conclude this section by noting that the system~\eqref{eq:free-oscillators} 
is an interpretation of the attention mechanism as a non-interacting spin 
glass: \eqref{eq:free-oscillators} is the projected gradient dynamics 
of the $d$-vector Hamiltonian~\eqref{eq:d-vector-hamiltonian} with $J_{ij} = 0$, 
and $h_i$ given by~\eqref{eq:attention-field}. This motivates extensions where the
free oscillators do interact with one another,
\begin{equation}
  \label{eq:coupled-attention-oscillators}
  \dot{z}_i = (I - z_iz_i^\top)\left(\sum_{j=1}^{T}J_{ij}z_j + \sum_{j=1}^{T}W_{ij}r_j\right)
\end{equation} 
In the case where the coupling between nodes~$i$ and~$j$ 
is ferromagnetic, i.e. $J_{ij} > 0$, the degree to which the output token $y_k$ 
attends to the input token $x_i$ is correlated with the degree it
attends to the input token $x_j$, 
whereas if $J_{ij} < 0$, the levels of attention are anticorrelated. 
Such models of \emph{cooperative attention} were studied
from a statistical physics perspective in~\cite{GK-DKP:26}, where the sampling 
was performed using a Boltzmann machine rather than an oscillator network. 
It was found that adding the couplings between spins actually improves 
performance over the classical softmax attention, which suggests that the
oscillator-based model of cooperative attention~\eqref{eq:coupled-attention-oscillators} 
may achieve similar performance gains. 

\section{Conclusion}
In this survey we reviewed a paradigm of computation where 
information is encoded through the phases of oscillations, and computation
proceeds via the dynamics of a coupled oscillator network. This approach 
forms an attractive alternative to digital computing since (1) oscillator 
networks admit a variety of physical realizations, many of which are more 
efficient than digital electronics and (2) analog computers can solve 
problems that are intractable on digital computers. The approach we develop 
unites the dynamics of coupled oscillator networks, 
statistical mechanical models of disorder in spin glasses, computational 
neuroscience, and theoretical computer science into a single common framework. 
Within this framework, we demonstrated how oscillator networks 
may implement three computational primitives: solving combinatorial optimization problems, 
storage and retrieval of memories, and the attention mechanism, which we 
hope may form the basis of oscillator-based realizations of efficient and 
intelligent devices. 


\section{Author Information}

\begin{IEEEbiography}{Ahmed Allibhoy}{\,}(\texttt{aallibho@uci.edu}) 
Ahmed Allibhoy is currently a postdoctoral scholar at UC Irvine. 
He received the B.S. degree in math and
electrical engineering from UCLA, and the M.S.
and Ph.D. degrees in mechanical engineering
from UC San Diego. His main research
interests include online optimization of dynamical
systems, nonsmooth dynamics, synchronization in coupled oscillator networks, 
and their applications to network neuroscience, systems biology, and artificial
intelligence.
\end{IEEEbiography}

\begin{IEEEbiography}{Fabio Pasqualetti}{\,}(\texttt{fabiopas@uci.edu}) 
Fabio Pasqualetti is a Professor of Electrical Engineering and Computer Science at UC Irvine, 
where he directs the UCI Robotarium. His research sits at the intersection of dynamical systems, 
machine learning, and computational neuroscience. He received the B.S. degree in 
computer engineering, the M.S. degree in automation engineering from the University of Pisa, and 
the Ph.D. degree in mechanical engineering from UC Santa Barbara. 
\end{IEEEbiography}

\bibliographystyle{IEEEtran}
\bibliography{refs}

\endarticle

\end{document}